\documentclass[10pt]{article}
\usepackage[latin9]{inputenc}
\usepackage{color}
\usepackage{mathrsfs}
\usepackage{amsmath}
\usepackage{amssymb}
\usepackage{graphicx}
\usepackage{setspace}
\usepackage[unicode=true,pdfusetitle,
 bookmarks=true,bookmarksnumbered=false,bookmarksopen=false,
 breaklinks=false,pdfborder={0 0 1},backref=false,colorlinks=false]
 {hyperref}

\makeatletter

\providecommand{\tabularnewline}{\\}

\@ifundefined{date}{}{\date{}}
\pdfoutput=1

\usepackage{cite}

\makeatother

\begin{document}
\noindent \begin{flushleft}
\textbf{\Large{}The complexity of dynamics in small neural circuits}\\

\par\end{flushleft}{\Large \par}

\noindent \begin{flushleft}
Diego Fasoli$^{1,\ast}$, Anna Cattani$^{1}$, Stefano Panzeri$^{1}$ 
\par\end{flushleft}

\medskip{}

\noindent \begin{flushleft}
\textbf{{1} Neural Computation Laboratory, Center for Neuroscience
and Cognitive Systems @Unitn, Istituto Italiano di Tecnologia, 38068
Rovereto (Tn), Italy }
\par\end{flushleft}

\noindent \begin{flushleft}
\textbf{$\ast$ E-mail: diego.fasoli@iit.it}
\par\end{flushleft}

\section*{Abstract}

\noindent Mean-field theory is a powerful tool for studying large
neural networks. However, when the system is composed of a few neurons,
macroscopic differences between the mean-field approximation and the
real behavior of the network can arise. Here we introduce a study
of the dynamics of a small firing-rate network with excitatory and
inhibitory populations, in terms of local and global bifurcations
of the neural activity. Our approach is analytically tractable in
many respects, and sheds new light on the finite-size effects of the
system. In particular, we focus on the formation of multiple branching
solutions of the neural equations through spontaneous symmetry-breaking,
since this phenomenon increases considerably the complexity of the
dynamical behavior of the network. For these reasons, branching points
may reveal important mechanisms through which neurons interact and
process information, which are not accounted for by the mean-field
approximation.

\section*{Author Summary}

\noindent The mesoscopic scale represents the bridge between microscopic
neural activity and the highest cognitive functions that emerge at
the macroscopic level in the brain. For this reason, understanding
the dynamics of small neural networks at this intermediate scale of
organization is of fundamental importance. However, counter-intuitively,
the behavior of small neural networks can be much more difficult to
study mathematically than that of large networks. Here we introduce
a finite-size firing-rate model and we study its local and global
bifurcations by a combination of numerical and analytical techniques.
This analysis shows the formation of strong and previously unexplored
finite-size effects, that are particularly hard to detect in large
networks. Their study advances the state of the art in the comprehension
of neural circuits, beyond the results provided by the mean-field
approximation and the techniques introduced so far for the quantification
of finite-size effects.

\section{Introduction \label{sec:Introduction}}

\noindent The structural complexity of the brain is reflected by its
organization at multiple spatial scales \cite{Sporns2006}. At the
highest level, the brain can be divided in macroscopic areas containing
millions of neurons. These areas accomplish very complex roles, ranging
from motor control to cognitive functions. On the opposite side, namely
at the lowest level, the brain is made of its elementary blocks, the
neurons, which provide a description of this complex organ at the
microscopic scale. The level of spatial organization that lies between
these macroscopic and microscopic pictures of the brain is called
\textit{mesoscopic scale}, and in the last years an increasing number
of studies (e.g. \cite{Ingber1992,Freeman2000a,Freeman2000b,Wright2003,Bohland2009})
have recognized its importance for the comprehension of the brain.
This is the spatial scale that links the macroscopic and microscopic
scales, and therefore that may explain how the highest cognitive functions
arise from the cooperation and the exchange of information between
neurons. For this reason, finding an appropriate mathematical description
of the brain at the mesoscopic scale is of fundamental importance
for unveiling its emergent properties.

At the mesoscopic scale, the brain is often described as a collection
of neural masses, i.e. homogeneous neuronal populations within a cortical
column \cite{Grimbert2008}. Usually, these groups of neurons are
described by the so called \textsl{neural-mass models} \cite{Deco2008}.
A typical example is the well-known \textsl{Jansen-Rit model} \cite{Jansen1993,Jansen1995,Babajani2006},
which describes a cortical minicolumn by a population of excitatory
pyramidal cells, that receive inhibitory and excitatory feedback from
local interneurons, and excitatory input from neighboring areas such
as the thalamus. This model was originally intended to describe the
formation of alpha rhythms within a cortical minicolumn of the visual
cortex through the interaction of the excitatory and inhibitory populations,
but provides only a coarse description of their dynamics. In \cite{Faugeras2009}
the authors proposed an improvement of the Jansen-Rit model based
on the mean-field theory, which may provide a better understanding
of MEG/EEG cortical recordings. The mean-field theory is a mathematical
tool that approximates the behavior of large networks \cite{Samuelides2007,Touboul2012,Baladron2012},
and proves very convenient from the computational point of view since
it describes the activity of large populations by means of a reduced
number of equations, which are usually simpler to solve than the original
network. This approximation becomes exact in the limit of networks
with an infinite number of neurons, the so called \textit{thermodynamic
limit}. Clearly this means that for finite-size networks, the mean-field
theory provides only an approximation of the real behavior of the
system, and therefore may neglect important phenomena.

For example, in \cite{Cessac1995}, for a class of random neural networks
it was proved that in the thermodynamic limit the system undergoes
a sharp transition from a static regime to chaos, while in the finite-size
case this occurs gradually by a cascade of transitions that are not
displayed by the mean-field approximation. Another important example
is shown in \cite{Fasoli2015}, where the authors proved that a stochastic
finite-size network can exhibit arbitrarily high levels of cross-correlations
of the neural activity, while this phenomenon disappears when the
size of the system grows to infinity. Clearly, these macroscopic differences
in the dynamical and statistical behavior of finite and infinite-size
networks may have important consequences on the information processing
capability of the system. In other terms, the oversimplification of
the mean-field approximation may hide important neural processes that
are fundamental for the comprehension of the brain.

This explains the number of mathematical techniques that have been
developed to quantify finite-size effects beyond the mean-field theory,
such as the linear noise approximation \cite{Bressloff2010}, the
density functional approach \cite{Buice2013}, large-deviations theory
\cite{Faugeras2013}, path-integral methods \cite{Bressloff2015},
etc. Typically, these finite-size methods can be applied to networks
composed of a finite but large number of neurons. However, neural
masses in a cortical minicolumn contain only few tens of neurons since,
according to different estimates \cite{Mountcastle1997,Buxhoeveden2002,Sporns2005},
a minicolumn in primates contains about $80-100$ neurons. Thus, for
such small networks the mean-field description and the previously
introduced techniques turn out to be inappropriate, since their finite-size
effects can be relevant.

The analysis of the dynamics of small neural networks was started
by Beer \cite{Beer1995}, who studied the bifurcations of networks
of arbitrary size in highly symmetric cases, namely with rigid constraints
on the strength of the synaptic weights. Through this analysis, he
showed that small networks can exhibit complex dynamical behavior,
which may have important neurobiological implications. In our article
we extend his analysis to a more biologically plausible network of
arbitrary size without specific conditions on the synaptic weights,
and in particular we show another important difference between a finite-size
network and its mean-field counterpart. In more detail, here we consider
a deterministic finite-size firing-rate network model with excitatory
and inhibitory populations composed of arbitrarily few neurons (we
consider $N_{E}$ neurons in the excitatory population and $N_{I}$
neurons in the inhibitory one), which are indistinguishable within
each population. Then, we perform a numerical analysis of the dynamics
that emerge by varying the external input current to the network and
the strength of inhibition, and we introduce a mathematical theory
that allows us to describe local bifurcations analytically. In particular,
we obtained macroscopic differences with the mean-field approximation
when the system has strong inhibitory synaptic weights. In this case,
through a phenomenon of spontaneous symmetry-breaking, the neural
network undergoes a special bifurcation known as \textit{branching
point} \cite{Kuznetsov1998elements}, from which multiple solutions
of the neural equations emerge. On the new branches, new bifurcations
can occur, enriching considerably the complexity of the bifurcation
diagram of the neural network. For this reason, symmetry and branching
points bifurcations may reveal important mechanisms through which
neurons interact and process information, which are not accounted
for by the mean-field approximation. Therefore a detailed analysis
of their properties may be very important in order to understand the
role of the mesoscopic scale in the emergence of brain's highest functions.

It is important to observe that since we consider identical neurons
within each population, our model lies in the context of the bifurcation
theory of dynamical systems with symmetry, which is known as \textit{equivariant
bifurcation theory} \cite{Golubitsky2012}. This usually requires
the reader to be familiar with advanced notions of group theory. Notwithstanding,
here we follow a simpler approach, so the mathematical analysis introduced
in this article is self-contained and accessible also to non-technical
readers.

Symmetry-breaking and branching point bifurcations have already been
studied in neuroscience through equivariant bifurcation theory, but
in a conceptually different way. For example, in the theory of visual
hallucinations developed by Ermentrout and Cowan \cite{Ermentrout1979},
symmetry was used to evaluate hallucinations in primary visual cortex
under the effect of drugs. They idealized the cortex as a plane and
described the local activity of a population of neurons by means of
neural field equations, so that each population is univocally identified
by its position in the continuous space of the plane. Their theory
exploits the symmetry the neural field equations inherit from the
geometrically regular structure of the anatomical connections in primary
visual cortex. In equivariant bifurcation theory, this symmetry is
described by the invariance of the equations under the action of the
Euclidean group $\boldsymbol{\mathrm{E}}\left(2\right)$, which is
the group of rigid motions in the plane, generated by translations,
rotations, and reflections.

However, the network analyzed in our work is made of two populations
containing a finite number of neurons, which are identified by a discrete
index. In each population the neurons are identical, therefore the
equations are symmetric under permutations of the neural indexes within
a population. In more mathematical terms, our equations are invariant
under the action of the group $S_{N_{E}}\times S_{N_{I}}$, where
$S_{N_{\alpha}}$ is the permutation group on $N_{\alpha}$ items
(also known as \textit{symmetric group}). This is a completely different
kind of symmetry compared to that of the Euclidean group $\boldsymbol{\mathrm{E}}\left(2\right)$,
and it allows us to study in an analytically simple way networks made
of a finite number of neurons. Indeed, this is conceptually different
from the theory of Ermentrout and Cowan, which relies on infinite-dimensional
neural field equations. So their theory does not describe finite-size
effects, and does not use the underlying symmetry for this purpose.
In this respect, our theory is more similar to that developed in \cite{Cohen2000,Stewart2003,Dias2003,Elmhirst2004},
where the authors exploited the invariance under the symmetric group
$S_{N}$ in a system with all-to-all coupling, and the subsequent
spontaneous symmetry-breaking, as a possible mechanism to explain
the origin of animal species. However, we are not aware of any application
of this symmetry to the study of bifurcations in neural networks described
by specific equations, such as Hopfield \cite{Hopfield1984} or Wilson-Cowan
\cite{WilsonCowan1972} ones.

This article is organized as follows. In Sec.~\eqref{sec:Materials-and-Methods}
we describe the neural model we use. Then we start Sec.~\eqref{sec:Results}
by explaining intuitively the main topic of this article, namely the
formation of new branches of solutions through spontaneous symmetry-breaking,
depending on the strength of inhibition (see SubSec.~\eqref{sub:Intuitive-interpretation-of-the-branching-points}).
This is followed by a numerical and analytical study of the bifurcation
points of the network in weak and strong-inhibition regimes (see SubSecs.~\eqref{sub:Weak-inhibition-regime}
and \eqref{sub:Strong-inhibition-regime} respectively). As we will
see, the complexity of dynamics depends also on the size of the inhibitory
population, therefore we start our analysis by showing the behavior
of the network in the simplest case, namely $N_{I}=2$, and then we
explain how to generalize these results to a larger inhibitory population
(see SubSec.~\eqref{sub:The-case-with-generic-NI}). We also show
why the mean-field theory is an oversimplified description of the
network (SubSec.~\eqref{sub:Differences-between-our-approach-and-the-mean-field-theory}),
and that the finite size effects studied in this work for a fully-connected
network may be even stronger for more realistic topologies of the
synaptic connections (SubSec.~\eqref{sub:Finite-size-effects-are-stronger-for-biologically-plausible-anatomical-connections}).
In Sec.~\eqref{sec:Discussion} we examine the importance and the
biological implications of our results. Finally, more details about
the analytical calculations are shown in Supplementary Materials.

\section{Materials and Methods \label{sec:Materials-and-Methods}}

\noindent The analysis of the dynamics of a neural network is a mathematically
hard problem. Its complexity depends on the degree of biological plausibility
of the system, which in turn is determined by:
\begin{itemize}
\item the variety of neural populations and realistic ratios between their
sizes;
\item the finite size of the network;
\item the random behavior of the system;
\item a realistic topology of the synaptic connections;
\item plausible single-cell models for the soma and the synapse;
\item delays in the transmission of the electric signals through the axons.
\end{itemize}
\noindent Among the points listed above, we select several biological
plausible features that lead our model to be characterized by:
\begin{itemize}
\item two neural populations of excitatory and inhibitory neurons;
\item a generic finite number of neurons in the network and in each population;
\item non-symmetric and arbitrarily strong synaptic weights.
\end{itemize}
\noindent For the sake of clarity, in this article we perform a numerical
and analytical analysis in the case of just two neural populations,
since they are sufficient for describing dynamical behaviors, such
as neural oscillations, which are thought to be at the base of fundamental
cognitive processes. It is important to observe that our analysis
can be extended to a generic number of neural populations, as we discuss
briefly in SubSec.~(S4.1) of the Supplementary Materials, but this
problem will be addressed in more detail in another article. In order
to make the analysis of the two-populations network analytically tractable,
we make some simplifying assumptions that considerably reduce the
complexity of the problem, without losing the most fundamental properties
of a neural network:
\begin{itemize}
\item non-spiking neurons;
\item identical neurons in each population;
\item dense connections (i.e. fully-connected topology);
\item absence of axonic delays;
\item absence of random noise.
\end{itemize}
\noindent These features are supposed to not compromise the plausibility
of the final result. Actually, this work is intended as a model of
a neural mass within a cortical column, where neurons are homogeneous
\cite{Grimbert2008} and the density of the connections is known to
be high, according to anatomical data (see for example \cite{Sporns2006}).
This justifies the use of identical neurons in each populations and
of a fully-connected topology. Moreover, since a single column is
localized in a mesoscopic area of the brain, delays can be neglected.
However, following \cite{Engelborghs2002,Yi2006}, our analysis can
be extended to the case of delay differential equations, if desired.
Also random noise may be taken into account, as shown in \cite{Fasoli2015},
but this will be studied in detail in another article. To conclude,
the choice of spiking or rate models is still under debate in the
scientific community (see for example \cite{Shadlen1994,Gerstner1997,Salinas2000,Panzeri2010})
so this, together with the need for a mathematically tractable model,
justifies our decision to work with a rate neural network. 

In more detail, we consider a widely used model of rate network \cite{Hopfield1984,Beer1995,Grimbert2008,Faugeras2009,Touboul2012,Fasoli2015}:

\begin{spacing}{0.8}
\begin{center}
{\small{}
\begin{equation}
\frac{dV_{i}\left(t\right)}{dt}=-\frac{1}{\tau_{i}}V_{i}\left(t\right)+\frac{1}{M_{i}}\sum_{j=0}^{N-1}J_{ij}\mathscr{A}_{j}\left(V_{j}\left(t\right)\right)+I_{i},\quad i=0,...,N-1\label{eq:exact-rate-equations-1}
\end{equation}
}
\par\end{center}{\small \par}
\end{spacing}

\noindent \begin{flushleft}
where $N\geq2$ represents the number of neurons in the network. The
function $V_{i}\left(t\right)$ is the membrane potential of the $i$th
neuron, while $\tau_{i}$ is its membrane time constant. $M_{i}$,
the total number of connections to the $i$th neuron, is a normalization
factor that has been introduced to avoid the explosion of the total
synaptic input $\sum_{j=0}^{N-1}J_{ij}\mathscr{A}_{j}\left(V_{j}\left(t\right)\right)$
in equation \eqref{eq:exact-rate-equations-1} in the thermodynamic
limit $N\rightarrow\infty$. The matrix $J$, also known as \textit{structural}
or \textit{anatomical connectivity}, represents the specification
of all the synaptic wirings that are physically present between neurons.
It also quantifies the strength of these connections. So $J_{ij}$
is the synaptic weight from the $j$th to the $i$th neuron, and for
simplicity it is supposed to be deterministic and constant in time,
for every pair of neurons. $\mathscr{A}_{j}\left(\cdot\right)$ is
the activation function of the $j$th neuron and converts its membrane
potential into the corresponding firing rate $\nu_{j}=\mathscr{A}\left(V_{j}\right)$.
$S$-shaped (i.e. sigmoidal) activation functions are biologically
plausible, however usually piecewise-linear functions are used in
order to find analytical results \cite{Campbell1996,Hansel1998,Ledoux2011}.
In our work we show how it is possible to obtain analytical expressions
for the equilibrium points and the local bifurcations of the network
when $\mathscr{A}\left(V\right)$ is the so called \textit{algebraic
activation function}:
\par\end{flushleft}

\begin{spacing}{0.8}
\begin{center}
{\small{}
\begin{equation}
\mathscr{A}_{j}\left(V\right)=\frac{\nu_{j}^{\mathrm{max}}}{2}\left[1+\frac{\frac{\Lambda_{j}}{2}\left(V-V_{j}^{T}\right)}{\sqrt{1+\frac{\Lambda_{j}^{2}}{4}\left(V-V_{j}^{T}\right)^{2}}}\right]\label{eq:algebraic-activation-function}
\end{equation}
}
\par\end{center}{\small \par}
\end{spacing}

Here $\nu^{\mathrm{max}}$ is the maximum firing rate, $\Lambda$
determines the slope of the activation function when $\nu^{\mathrm{max}}$
is fixed, and $V^{T}$ is the horizontal shift, so it can be interpreted
as a firing threshold for the membrane potentials. In Eq.~\eqref{eq:exact-rate-equations-1},
$I_{i}$ are external input currents, which in this article are supposed
to be constant in time, in order to perform the bifurcation analysis
of the network.

In order to make our analysis analytically tractable, we suppose that
all the parameters of the system are indexed only at the population
level. In other terms, within a given population, or between two fixed
populations in the case of the synaptic weights, the parameters are
homogeneous. If we define $N_{E}$ ($N_{I}$) to be the size of the
excitatory (inhibitory) population, and if we suppose that the neurons
with indexes $i=0,...,N_{E}-1$ belong to the excitatory population,
and those with $i=N_{E},...,N-1$ (with $N=N_{E}+N_{I}$) to the inhibitory
one, this means that the synaptic connectivity matrix is structured
as follows:

\begin{spacing}{0.8}
\begin{center}
{\small{}
\[
\begin{array}{ccc}
J=\left[\begin{array}{cc}
\mathscr{\mathfrak{J}}_{EE} & \mathscr{\mathfrak{J}}_{EI}\\
\mathscr{\mathfrak{J}}_{IE} & \mathscr{\mathfrak{J}}_{II}
\end{array}\right], &  & \mathfrak{J}_{\alpha\beta}=\begin{cases}
J_{\alpha\alpha}\left(\mathbb{I}_{N_{\alpha}}-\mathrm{Id}_{N_{\alpha}}\right), & \;\mathrm{for}\;\alpha=\beta\\
\\
J_{\alpha\beta}\mathbb{I}_{N_{\alpha},N_{\beta}}, & \;\mathrm{for}\;\alpha\neq\beta
\end{cases}\end{array}
\]
}
\par\end{center}{\small \par}
\end{spacing}

\noindent \begin{flushleft}
where the block $\mathfrak{J}_{\alpha\beta}$ is a $N_{\alpha}\times N_{\beta}$
matrix that represents the connections from the population $\beta$
to the population $\alpha$, with $\alpha,\beta\in\left\{ E,I\right\} $.
Moreover, $\mathbb{I}_{N_{\alpha},N_{\beta}}$ is the $N_{\alpha}\times N_{\beta}$
all-ones matrix (here we use the simplified notation $\mathbb{I}_{N_{\alpha}}\overset{\mathrm{def}}{=}\mathbb{I}_{N_{\alpha},N_{\alpha}}$),
while $\mathrm{Id}_{N_{\alpha}}$ is the $N_{\alpha}\times N_{\alpha}$
identity matrix. Since, according to experimental measurements, about
$80\%$ of neocortical neurons are excitatory and the remaining $20\%$
are inhibitory \cite{Markram2004}, in this article we consider $\frac{N_{E}}{N_{I}}=4$.
The zeros on the diagonal lines of the matrices $\mathfrak{J}_{EE}$
and $\mathfrak{J}_{II}$ are due to the absence of self-connections
in biological networks. The real numbers $J_{EE}$, $J_{II}$, $J_{EI}$,
$J_{IE}$ are free parameters that describe the strength of the interactions
between and within the neural populations. Clearly we have $J_{EE},J_{IE}>0$,
and $J_{II},J_{EI}<0$, which also means that $M_{E}=M_{I}=N-1$.
It is important to observe that, compared to \cite{Touboul2012},
we have chosen a different normalization of the synaptic weights.
In both the cases, the total synaptic input $\sum_{j=0}^{N-1}J_{ij}\mathscr{A}_{j}\left(V_{j}\left(t\right)\right)$
is convergent in the thermodynamic limit, but they behave in different
ways when the network is mixed with both finite-size and infinite-size
populations. It is easy to check that according to our normalization,
if one population has infinite size, then the contribution of the
finite-size population in the total synaptic input goes to zero, which
seems reasonable to us. On the other side, with the normalization
introduced in \cite{Touboul2012}, the infinite-size and finite-size
populations both provide finite contributions to the total synaptic
input. However, if required, it is possible to switch from our normalization
to the other one by performing the following substitution:
\par\end{flushleft}

\begin{spacing}{0.8}
\begin{center}
{\small{}
\[
J_{\alpha\beta}\rightarrow\frac{N-1}{N_{\beta}}J_{\alpha\beta}
\]
}
\par\end{center}{\small \par}
\end{spacing}

\noindent in all the analytical results that we will obtain in the
next sections and in the Supplementary Materials.

\begin{figure}
\begin{centering}
\includegraphics[scale=0.33]{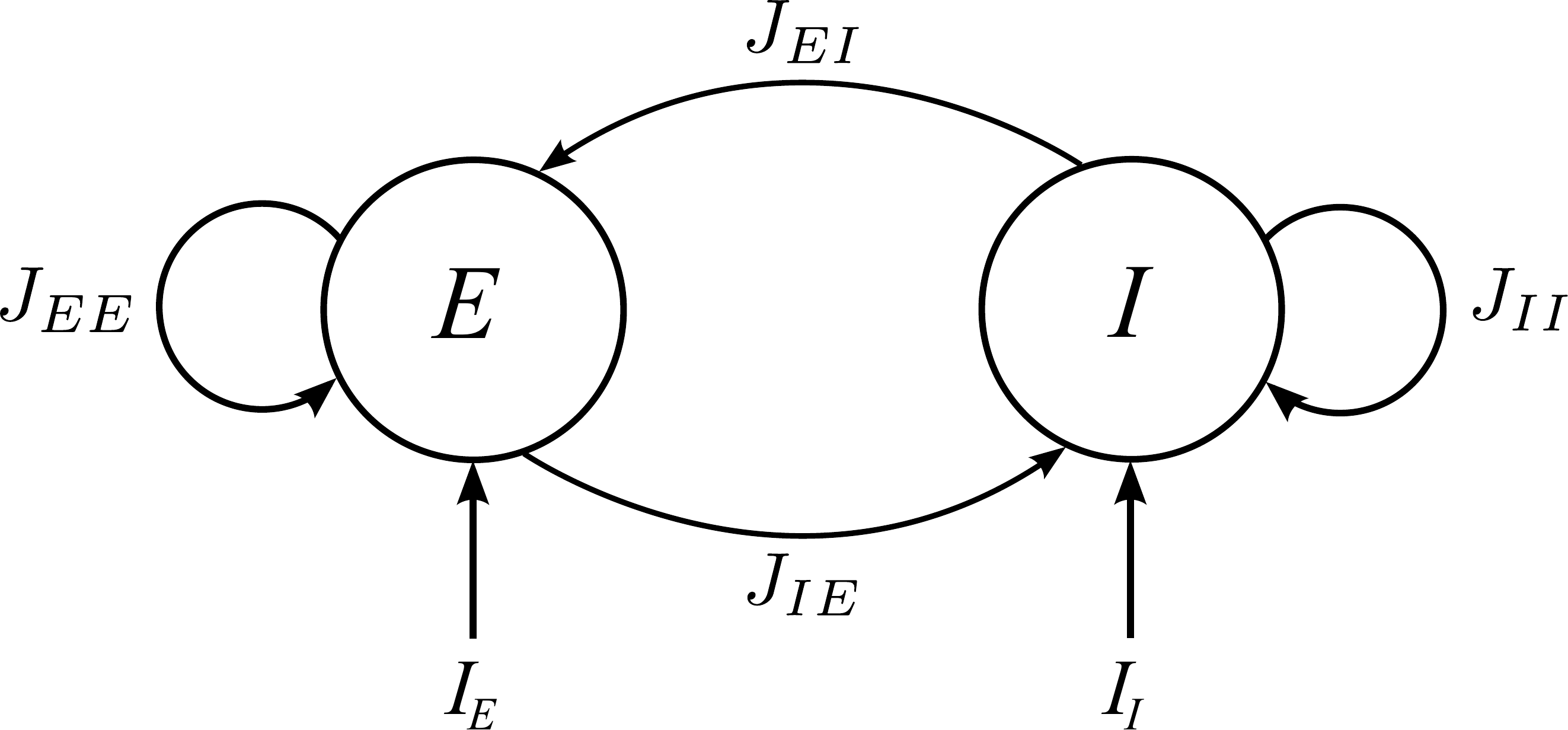}
\par\end{centering}

\protect\caption{\label{Fig:network-structure} \small Structure of the network considered
in this article.}
\end{figure}

Moreover, from our assumption on the indexes, we obtain that the external
input currents are divided in two vectors, $\boldsymbol{I}_{E}$ and
$\boldsymbol{I}_{I}$, namely the inputs to the excitatory and inhibitory
populations. Clearly we get:

\begin{spacing}{0.8}
\begin{center}
{\small{}
\[
\boldsymbol{I}_{\alpha}=I_{\alpha}\boldsymbol{1}_{N_{\alpha}}
\]
}
\par\end{center}{\small \par}
\end{spacing}

\noindent where $\boldsymbol{1}_{N_{\alpha}}$ is the $N_{\alpha}\times1$
all-ones vector. Therefore the structure of the network is that shown
in Fig.~\eqref{Fig:network-structure}. The same subdivision between
excitatory and inhibitory populations is performed for the other parameters
of the network, namely $\tau$, $\nu^{\mathrm{max}}$, $\Lambda$,
$V_{T}$. 

To conclude, for the sake of clarity we observe that under our assumption
on the neural indexes, we can rewrite the system \eqref{eq:exact-rate-equations-1}
in the following explicit way:

\begin{spacing}{0.8}
\begin{center}
{\scriptsize{}
\begin{equation}
\begin{cases}
E:\quad\frac{dV_{i}\left(t\right)}{dt}= & -\frac{1}{\tau_{E}}V_{i}\left(t\right)+\frac{J_{EE}}{N-1}{\displaystyle \sum_{\substack{j=0\\
j\neq i
}
}^{N_{E}-1}}\mathscr{A}_{E}\left(V_{j}\left(t\right)\right)+\frac{J_{EI}}{N-1}{\displaystyle \sum_{j=N_{E}}^{N-1}}\mathscr{A}_{I}\left(V_{j}\left(t\right)\right)+I_{E},\quad i=0,...,N_{E}-1\\
\\
I:\quad\frac{dV_{i}\left(t\right)}{dt}= & -\frac{1}{\tau_{I}}V_{i}\left(t\right)+\frac{J_{IE}}{N-1}{\displaystyle \sum_{j=0}^{N_{E}-1}}\mathscr{A}_{E}\left(V_{j}\left(t\right)\right)+\frac{J_{II}}{N-1}{\displaystyle \sum_{\substack{j=N_{E}\\
j\neq i
}
}^{N-1}}\mathscr{A}_{I}\left(V_{j}\left(t\right)\right)+I_{I},\quad i=N_{E},...,N-1
\end{cases}\label{eq:exact-rate-equation-2}
\end{equation}
}
\par\end{center}{\scriptsize \par}
\end{spacing}

\section{Results\label{sec:Results}}

\noindent The bifurcation analysis we perform provides an overview
of the dynamics the model is able to exhibit, depending on two parameters:
the static external currents $I_{E,I}$ in \eqref{eq:exact-rate-equation-2}.
In particular, we present this analysis for increasing values of a
third parameter: the self-inhibition strength $J_{II}$. We focus
on this parameter, instead of the other synaptic weights (i.e. $J_{EE}$,
$J_{EI}$, $J_{IE}$), since $J_{II}$ is directly responsible for
the main phenomenon analyzed in this article, namely the formation
of the branching point bifurcations.\textcolor{blue}{{} }For the remaining
synaptic weights we will provide only a qualitative description of
the effects they exert on the system. The non-varying network parameters
are chosen as in Tab.~\eqref{Tab:parameters}. 
\begin{table}
\begin{centering}
{\small{}}%
\begin{tabular}{|c|c|c|c|}
\hline 
\textbf{\small{}Population Size} & \textbf{\small{}Synaptic Weights} & \textbf{\small{}Activation Functions} & \textbf{\small{}Other}\tabularnewline
\hline 
{\small{}$N_{E}=8$} & {\small{}$J_{EE}=10$} & {\small{}$\nu_{E}^{\mathrm{max}}=\nu_{I}^{\mathrm{max}}=1$} & {\small{}$\tau_{E}=\tau_{I}=1$}\tabularnewline
\hline 
{\small{}$N_{I}=2$} & {\small{}$J_{EI}=-70$} & {\small{}$\Lambda_{E}=\Lambda_{I}=2$} & \tabularnewline
\hline 
 & {\small{}$J_{IE}=70$} & {\small{}$V_{E}^{T}=V_{I}^{T}=2$} & \tabularnewline
\hline 
\end{tabular}
\par\end{centering}{\small \par}

\protect\caption{\label{Tab:parameters} \small Values of the parameters used in this
article.}
\end{table}
 In particular, we consider a network made of $N_{E}=8$ excitatory
neurons and $N_{I}=2$ inhibitory ones, since we want to study finite-size
effects in small neural masses.

We perform a detailed bifurcation analysis by means of numerical tools
and, when possible, through analytical techniques. The numerical analysis
is performed by exploiting the Cl\_MatCont Matlab toolbox \cite{DhoogeGovaerts}
and XPPAUT \cite{Ermentrout2002}, that are grounded in the mathematical
theory of bifurcations described in \cite{Kuznetsov1998elements,Strogatz1994nonlinear},
while the analytical results are based on elementary methods from
linear algebra. It is important to underline that bifurcations are
defined by many conditions. Nonetheless, in our analytical study we
checked only the conditions on the eigenvalues of the network, since
they proved sufficient to reproduce the numerical results. Due to
the high variety of the bifurcations the system exhibits, a rigorous
check of all the remaining conditions is beyond the purpose of this
article, and is left to the most technical readers.

\subsection{Intuitive interpretation of the branching points \label{sub:Intuitive-interpretation-of-the-branching-points}}

In mathematics, the branching point bifurcations are described by
the so-called \textit{equivariant bifurcation theory} \cite{Golubitsky2012},
namely the study of bifurcations in symmetric systems. Being the latter
rather technical, here we prefer to follow a more intuitive approach
to the problem. So, first of all, we have to observe that according
to bifurcation theory, local bifurcations are calculated by means
of the eigenvalues of the Jacobian matrix of the network, evaluated
at the equilibrium points. Therefore, we start by setting $\frac{dV_{i}\left(t\right)}{dt}=0\;\forall i$
in Eq.~\eqref{eq:exact-rate-equation-2}. The main observation of
this article is that the system shows two different behaviors depending
on the strength of $J_{II}$. As far as inhibition is weak (this will
be quantified rigorously below), the equilibrium points in each population
are homogeneous:

\begin{spacing}{0.8}
\begin{center}
{\small{}
\begin{equation}
\boldsymbol{\mu}=\left(\overset{N_{E}\mathrm{-times}}{\overbrace{\mu_{E},\ldots,\mu_{E}}},\overset{N_{I}\mathrm{-times}}{\overbrace{\mu_{I},\ldots,\mu_{I}}}\right)\label{eq:equilibrium-points-homogeneous-case}
\end{equation}
}
\par\end{center}{\small \par}
\end{spacing}

\noindent \begin{flushleft}
where $\mu_{E}$ and $\mu_{I}$ are the solutions of the following
system of algebraic non-linear equations, obtained from \eqref{eq:exact-rate-equation-2}:
\par\end{flushleft}

\begin{spacing}{0.8}
\begin{center}
{\small{}
\begin{equation}
\begin{cases}
\mathscr{F}\left(\mu_{E},\mu_{I}\right)\overset{\mathrm{def}}{=}-\frac{1}{\tau_{E}}\mu_{E}+\frac{N_{E}-1}{N-1}J_{EE}\mathscr{A}_{E}\left(\mu_{E}\right)+\frac{N_{I}}{N-1}J_{EI}\mathscr{A}_{I}\left(\mu_{I}\right)+I_{E}=0\\
\\
\mathscr{G}\left(\mu_{E},\mu_{I}\right)\overset{\mathrm{def}}{=}-\frac{1}{\tau_{I}}\mu_{I}+\frac{N_{E}}{N-1}J_{IE}\mathscr{A}_{E}\left(\mu_{E}\right)+\frac{N_{I}-1}{N-1}J_{II}\mathscr{A}_{I}\left(\mu_{I}\right)+I_{I}=0
\end{cases}\label{eq:equilibrium-points-equations-homogeneous-case}
\end{equation}
}
\par\end{center}{\small \par}
\end{spacing}

\begin{figure}
\begin{centering}
\includegraphics[scale=0.19]{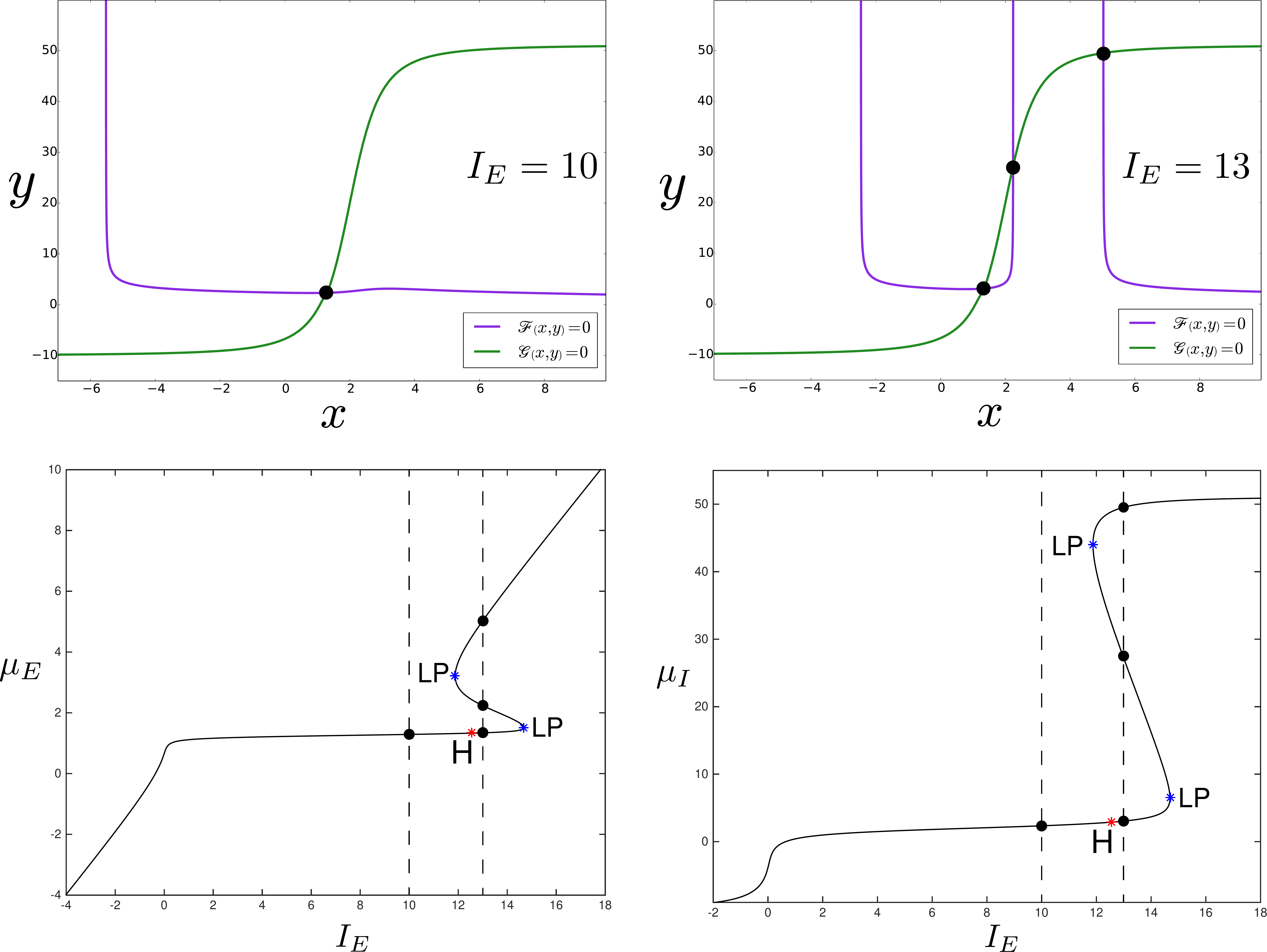}
\par\end{centering}

\protect\caption{\label{Fig:equilibrium-points-2D} \small The two figures on the
top show the nullclines of the network for fixed values of $I_{E,I}$
in a weak-inhibition regime, obtained for $J_{II}=-10$ and the values
of the parameters in Tab.~\eqref{Tab:parameters}. Their intersection
points (black dots) correspond to the solutions of the system \eqref{eq:equilibrium-points-equations-homogeneous-case}.
The top-left figure was obtained for $I_{E}=10$, $I_{I}=-10$, while
the top-right figure for $I_{E}=13$, $I_{I}=-10$. Clearly the system
\eqref{eq:equilibrium-points-equations-homogeneous-case} admits multiple
solutions for specific values of the parameters. The two figures at
the bottom show the solutions $\mu_{E}$ and $\mu_{I}$ (bottom-left
and bottom-right panel, respectively) of the system \eqref{eq:equilibrium-points-equations-homogeneous-case}
for the same values of the parameters, but with varying current $I_{E}$.
The black curve represents the primary branch of the network equations,
and for $I_{E}=10$ and $I_{E}=13$ it admits one and three solutions
respectively (see the black dots at the intersection with the vertical
dashed lines). The $\left(\mu_{E},\mu_{I}\right)$ coordinates of
these solutions correspond to those of the black dots in the top panels
of the figure. Here we do not care about the stability of the solutions,
which is shown for the sake of completeness in Fig.~(S5) of the Supplementary
Materials.}
 
\end{figure}

\noindent The curves defined by Eqs. $\mathscr{F}\left(x,y\right)=0$
and $\mathscr{G}\left(x,y\right)=0$ $\forall\left(x,y\right)\in\mathbb{R}^{2}$
are the so called \textit{nullclines} of the network. Fig.~\eqref{Fig:equilibrium-points-2D}
(top) shows an example obtained for $J_{II}=-10$, while the remaining
parameters are chosen as in Tab.~\eqref{Tab:parameters}. 
In Sec.~(S2) of the Supplementary Materials we show how to get approximate
analytical solutions for $\mu_{E,I}$.

From \eqref{eq:equilibrium-points-equations-homogeneous-case}, we
can see that the Jacobian matrix $\mathcal{J}$ of the network on
the primary branch of solutions \eqref{eq:equilibrium-points-homogeneous-case}
is:

\begin{spacing}{0.8}
\begin{center}
{\small{}
\begin{equation}
\begin{array}{ccc}
\mathcal{J}=\left[\begin{array}{cc}
\mathscr{J}_{EE} & \mathscr{J}_{EI}\\
\mathscr{J}_{IE} & \mathscr{J}_{II}
\end{array}\right], &  & \mathscr{J}_{\alpha\beta}=\begin{cases}
-\frac{1}{\tau_{\alpha}}\mathrm{Id}{}_{N_{\alpha}}+\frac{J_{\alpha\alpha}}{N-1}\mathscr{A}_{\alpha}'\left(\mu_{\alpha}\right)\left(\mathbb{I}_{N_{\alpha}}-\mathrm{Id}_{N_{\alpha}}\right), & \;\mathrm{for}\;\alpha=\beta\\
\\
\frac{J_{\alpha\beta}}{N-1}\mathscr{A}_{\beta}'\left(\mu_{\beta}\right)\mathbb{I}_{N_{\alpha},N_{\beta}}, & \;\mathrm{for}\;\alpha\neq\beta
\end{cases}\end{array}\label{eq:Jacobian-matrix}
\end{equation}
}
\par\end{center}{\small \par}
\end{spacing}

\noindent therefore we can prove (see SubSec.~(S3.1) of the Supplementary
Materials) that its eigenvalues are:

\begin{spacing}{0.8}
\begin{center}
{\small{}
\begin{equation}
\lambda_{0,1}=\frac{\delta+\eta\pm\sqrt{\left(\delta-\eta\right)^{2}+4\gamma}}{2},\quad\lambda_{E}=-\left[\frac{1}{\tau_{E}}+\frac{J_{EE}}{N-1}\mathscr{A}_{E}'\left(\mu_{E}\right)\right],\quad\lambda_{I}=-\left[\frac{1}{\tau_{I}}+\frac{J_{II}}{N-1}\mathscr{A}_{I}'\left(\mu_{I}\right)\right]\label{eq:eigenvalues-Jacobian-matrix}
\end{equation}
}
\par\end{center}{\small \par}
\end{spacing}

\noindent where:

\begin{spacing}{0.8}
\begin{center}
{\small{}
\begin{equation}
\delta=-\frac{1}{\tau_{E}}+\frac{N_{E}-1}{N-1}J_{EE}\mathscr{A}_{E}'\left(\mu_{E}\right),\quad\eta=-\frac{1}{\tau_{I}}+\frac{N_{I}-1}{N-1}J_{II}\mathscr{A}_{I}'\left(\mu_{I}\right),\quad\gamma=\frac{N_{E}N_{I}}{\left(N-1\right)^{2}}J_{EI}J_{IE}\mathscr{A}_{E}'\left(\mu_{E}\right)\mathscr{A}_{I}'\left(\mu_{I}\right)\label{eq:parameters-for-the-eigenvalues}
\end{equation}
}
\par\end{center}{\small \par}
\end{spacing}

\noindent According to bifurcation theory, the system undergoes special
bifurcations when one of its eigenvalues is equal to zero. In particular,
the branching point bifurcations are given by the condition $\lambda_{I}=0$,
so this allows us to define the weak and strong-inhibition regimes
quantitatively by the conditions $\lambda_{I}<0$ and $\lambda_{I}\geq0$,
respectively.

\begin{figure}
\begin{centering}
\includegraphics[scale=0.3]{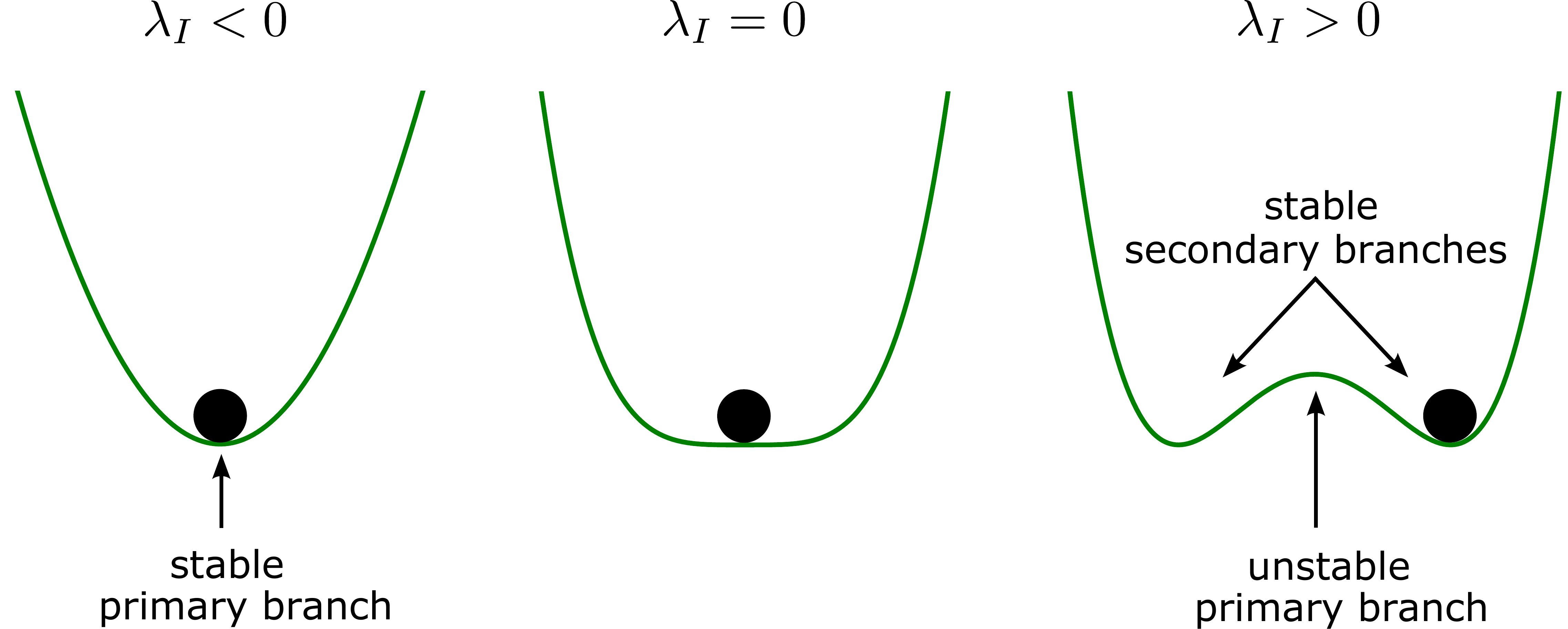}
\par\end{centering}

\protect\caption{{\small{}\label{Fig:symmetry-breaking} }\small Spontaneous symmetry-breaking.
For weak inhibition, the eigenvalue $\lambda_{I}$ (see Eq.~\eqref{eq:eigenvalues-Jacobian-matrix})
is negative, and the system \eqref{eq:exact-rate-equation-2} has
only one stationary solution. This solution is stable and symmetric,
but for increasing inhibition $\lambda_{I}$ changes sign, therefore
the solution becomes unstable and the network chooses randomly between
two new alternative stable states, breaking the symmetry. This phenomenon
can be understood intuitively by means of a ball that rolls in a double
well potential in order to reach a state of minimum energy, that we
interpret as a stable stationary solution of Eq.~\eqref{eq:exact-rate-equation-2}.}
\end{figure}

When the network undergoes a branching point bifurcation, we observe
the formation of heterogeneous membrane potentials in the inhibitory
population: in other terms, under strong inhibition the symmetry of
the system is broken. Intuitively, this can be understood as in Fig.~\eqref{Fig:symmetry-breaking}.
When $\left|J_{II}\right|$ is small, there is only one valley or
basin in the ``energy landscape'' of the network. On the other side,
for strong inhibition we observe the formation of multiple valleys,
and a small perturbation determines to which one the inhibitory potential
will converge. For this reason, now multiple new branches of the equilibrium
points emerge, which are described by the following stationary solutions:

\begin{spacing}{0.8}
\begin{center}
{\small{}
\begin{equation}
\boldsymbol{\mu}=\left(\overset{N_{E}\mathrm{-times}}{\overbrace{\mu_{E},\ldots,\mu_{E}}},\mu_{I,0},\ldots,\mu_{I,N_{I}-1}\right)\label{eq:equilibrium-points-heterogeneous-case}
\end{equation}
}
\par\end{center}{\small \par}
\end{spacing}

\noindent \begin{flushleft}
where $\mu_{E}$ and $\mu_{I,i}$ are the solutions of the following
system of algebraic non-linear equations, obtained from \eqref{eq:exact-rate-equation-2}
in the stationary regime:
\par\end{flushleft}

\begin{spacing}{0.8}
\begin{center}
{\small{}
\begin{equation}
\begin{cases}
-\frac{1}{\tau_{E}}\mu_{E}+\frac{N_{E}-1}{N-1}J_{EE}\mathscr{A}_{E}\left(\mu_{E}\right)+\frac{J_{EI}}{N-1}{\displaystyle \sum_{j=0}^{N_{I}-1}}\mathscr{A}_{I}\left(\mu_{I,j}\right)+I_{E}=0\\
\\
-\frac{1}{\tau_{I}}\mu_{I,i}+\frac{N_{E}}{N-1}J_{IE}\mathscr{A}_{E}\left(\mu_{E}\right)+\frac{J_{II}}{N-1}{\displaystyle \sum_{\substack{j=0\\
j\neq i
}
}^{N_{I}-1}}\mathscr{A}_{I}\left(\mu_{I,j}\right)+I_{I}=0 & \mathrm{for}\;i=0,\ldots,N_{I}-1
\end{cases}\label{eq:equilibrium-points-equations-heterogeneous-case-0}
\end{equation}
}
\par\end{center}{\small \par}
\end{spacing}

\noindent For example, for $N_{I}=2$ Eq.~\eqref{eq:equilibrium-points-equations-heterogeneous-case-0}
can be written more explicitly as follows:

\begin{spacing}{0.8}
\begin{center}
{\small{}
\begin{equation}
\begin{cases}
\mathscr{F}\left(\mu_{E},\mu_{I,0},\mu_{I,1}\right)\overset{\mathrm{def}}{=}-\frac{1}{\tau_{E}}\mu_{E}+\frac{N_{E}-1}{N-1}J_{EE}\mathscr{A}_{E}\left(\mu_{E}\right)+\frac{J_{EI}}{N-1}\left[\mathscr{A}_{I}\left(\mu_{I,0}\right)+\mathscr{A}_{I}\left(\mu_{I,1}\right)\right]+I_{E}=0\\
\\
\mathscr{G}\left(\mu_{E},\mu_{I,0},\mu_{I,1}\right)\overset{\mathrm{def}}{=}-\frac{1}{\tau_{I}}\mu_{I,0}+\frac{N_{E}}{N-1}J_{IE}\mathscr{A}_{E}\left(\mu_{E}\right)+\frac{J_{II}}{N-1}\mathscr{A}_{I}\left(\mu_{I,1}\right)+I_{I}=0\\
\\
\mathscr{H}\left(\mu_{E},\mu_{I,0},\mu_{I,1}\right)\overset{\mathrm{def}}{=}-\frac{1}{\tau_{I}}\mu_{I,1}+\frac{N_{E}}{N-1}J_{IE}\mathscr{A}_{E}\left(\mu_{E}\right)+\frac{J_{II}}{N-1}\mathscr{A}_{I}\left(\mu_{I,0}\right)+I_{I}=0
\end{cases}\label{eq:equilibrium-points-equations-heterogeneous-case-1}
\end{equation}
}
\par\end{center}{\small \par}
\end{spacing}

\begin{figure}
\begin{centering}
\includegraphics[scale=0.3]{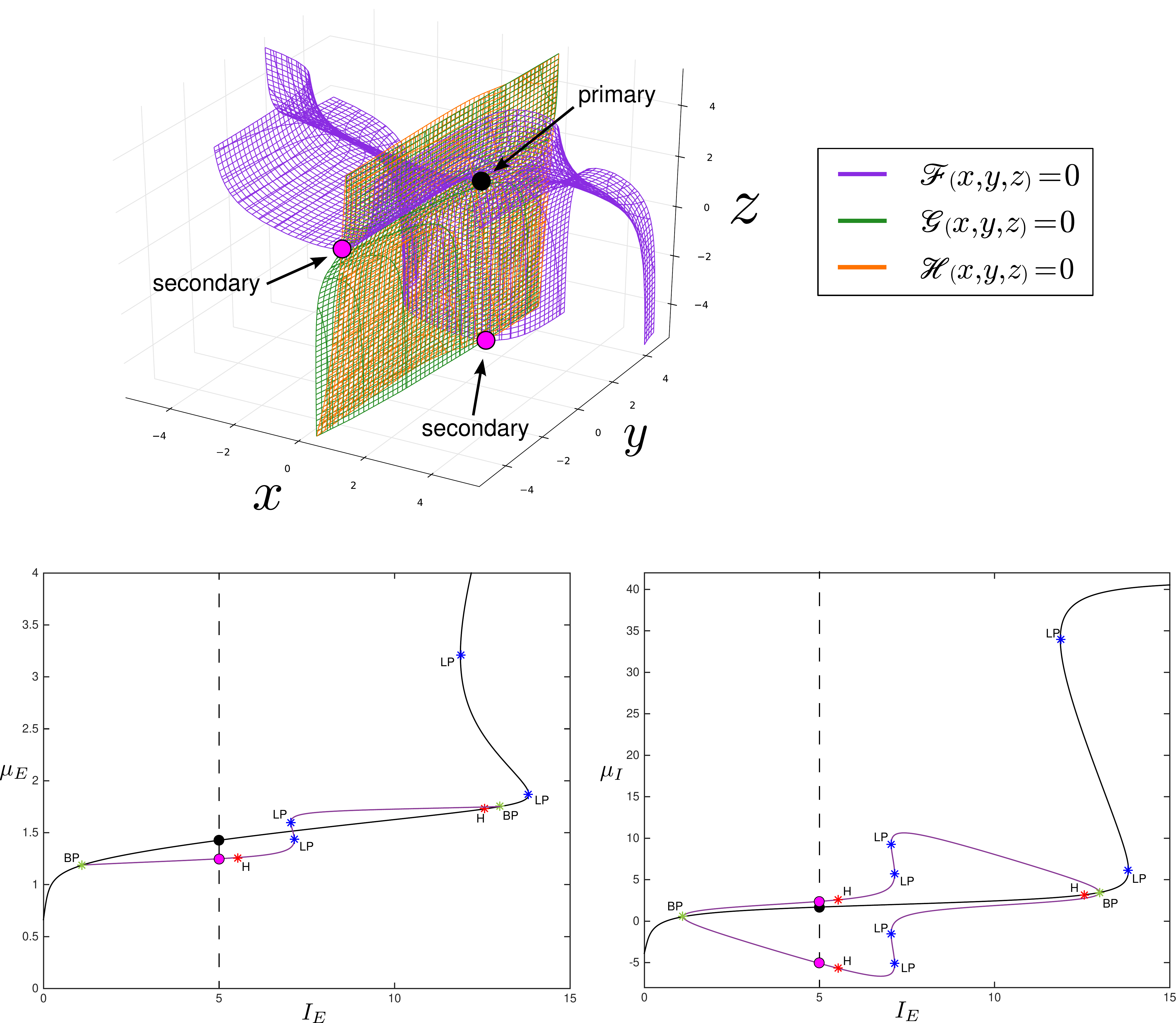}
\par\end{centering}

\protect\caption{\label{Fig:equilibrium-points-3D} \small The figure on the top shows
the nullsurfaces of the network for fixed values of $I_{E,I}$ in
a strong-inhibition regime, obtained for $J_{II}=-100$, $I_{E}=5$,
$I_{I}=-10$ and the values of the parameters in Tab.~\eqref{Tab:parameters}.
The black intersection point corresponds to the solution of the system
\eqref{eq:equilibrium-points-equations-heterogeneous-case-1} on the
primary branch, while the purple dots represent the solutions on the
secondary branches. The two figures at the bottom show the solutions
$\mu_{E}$ and $\mu_{I}$ (left and right panel, respectively) of
the system \eqref{eq:equilibrium-points-equations-heterogeneous-case-1}
for the same values of the parameters, but with varying current $I_{E}$.
The black and violet curves represent respectively the primary and
secondary branches of the network equations. For $I_{E}=5$ the system
admits three solutions $\left(\mu_{E},\mu_{I}\right)$ (see the dots
at the intersection with the vertical dashed lines: we have just three
solutions because $\mu_{I}$ has three intersection points, two of
which, i.e. the pink ones in the right panel, correspond to the same
$\mu_{E}$, i.e. the pink dot in the left panel). The $\left(\mu_{E},\mu_{I}\right)$
coordinates of these solutions correspond to those of the dots in
the top panel of the figure. Again, here we do not care about the
stability of the solutions, which is shown for the sake of completeness
in Fig.~\eqref{Fig:strong-inhibition-regime-1} and in Fig.~(S9)
of the Supplementary Materials.}
 
\end{figure}

\noindent The surfaces $\mathscr{F}\left(x,y,z\right)=0$, $\mathscr{G}\left(x,y,z\right)=0$,
$\mathscr{H}\left(x,y,z\right)=0$ $\forall\left(x,y,z\right)\in\mathbb{R}^{3}$
are a higher-dimensional extension of the nullclines $\mathscr{F}\left(x,y\right)=0$
and $\mathscr{G}\left(x,y\right)=0$ $\forall\left(x,y\right)\in\mathbb{R}^{2}$
that we encountered in the weak-inhibition regime. Sometimes they
are called \textit{nullsurfaces} (see for example \cite{Beer1995}).
Fig.~\eqref{Fig:equilibrium-points-3D} (top) shows an example obtained
for $J_{II}=-100$, while the remaining parameters are chosen as in
Tab.~\eqref{Tab:parameters}. 

For the sake of clarity, here we treat in detail only the case $N_{I}=2$,
while we will show some results for $N_{I}>2$ in SubSec.~\eqref{sub:The-case-with-generic-NI}.
So, for $N_{I}=2$, from Eq.~\eqref{eq:equilibrium-points-equations-heterogeneous-case-1}
we get that the Jacobian matrix on the secondary branches \eqref{eq:equilibrium-points-heterogeneous-case}
is:

\begin{spacing}{0.8}
\begin{center}
{\small{}
\begin{equation}
\mathcal{J}=\left[\begin{array}{ccc}
\mathscr{J}_{00} & \mathscr{J}_{01} & \mathscr{J}_{02}\\
\mathscr{J}_{10} & \mathscr{J}_{11} & \mathscr{J}_{12}\\
\mathscr{J}_{20} & \mathscr{J}_{21} & \mathscr{J}_{22}
\end{array}\right]\label{eq:Jacobian-matrix-on-the-new-branches}
\end{equation}
}
\par\end{center}{\small \par}
\end{spacing}

\noindent where:

\begin{spacing}{0.8}
\begin{center}
{\small{}
\begin{align*}
 & \mathscr{J}_{00}=-\frac{1}{\tau_{E}}\mathrm{Id}{}_{N_{E}}+\frac{J_{EE}}{N-1}\mathscr{A}_{E}'\left(\mu_{E}\right)\left(\mathbb{I}_{N_{E}}-\mathrm{Id}_{N_{E}}\right),\quad\mathscr{J}_{01}=\frac{J_{EI}}{N-1}\mathscr{A}_{I}'\left(\mu_{I,0}\right)\boldsymbol{1}_{N_{E}},\quad\mathscr{J}_{02}=\frac{J_{EI}}{N-1}\mathscr{A}_{I}'\left(\mu_{I,1}\right)\boldsymbol{1}_{N_{E}},\\
\\
 & \mathscr{J}_{10}=\frac{J_{IE}}{N-1}\mathscr{A}_{E}'\left(\mu_{E}\right)\boldsymbol{1}_{N_{E}}^{T},\quad\mathscr{J}_{11}=\frac{1}{\tau_{I}},\quad\mathscr{J}_{12}=\frac{J_{II}}{N-1}\mathscr{A}_{I}'\left(\mu_{I,1}\right),\\
\\
 & \mathscr{J}_{20}=\frac{J_{IE}}{N-1}\mathscr{A}_{E}'\left(\mu_{E}\right)\boldsymbol{1}_{N_{E}}^{T},\quad\mathscr{J}_{21}=\frac{J_{II}}{N-1}\mathscr{A}_{I}'\left(\mu_{I,0}\right),\quad\mathscr{J}_{22}=\frac{1}{\tau_{I}}
\end{align*}
}
\par\end{center}{\small \par}
\end{spacing}

\noindent (here $T$ is the transpose of a matrix), as explained in
more detail in SubSec.~(S4.1) of the Supplementary Materials. Intuitively,
on the primary branch the Jacobian matrix was a $2\times2$ block
matrix (see Eq.~\eqref{eq:Jacobian-matrix}), because we had only
an excitatory and an inhibitory membrane potential ($\mu_{E}$ and
$\mu_{I}$ respectively), while on the secondary branch it is a $3\times3$
block matrix, because now the two inhibitory neurons have different
potentials ($\mu_{I,0}$ and $\mu_{I,1}$). The Jacobian matrix on
the new branches can be calculated for a network with a generic number
of inhibitory neurons (see the Supplementary Materials for more details).

Finally, we observe that it is possible to find relations between
the inhibitory membrane potentials in the strong-inhibition regime,
which prove very useful when we calculate analytically the local bifurcations
of the system. For example, in the case $N_{I}=2$, from the second
and third equation of the system \eqref{eq:equilibrium-points-equations-heterogeneous-case-1},
after some algebra we get the following forth-order polynomial equation:

\begin{spacing}{0.8}
\begin{center}
{\small{}
\begin{equation}
\widehat{a}\mu_{I,1}^{4}+\widehat{b}\mu_{I,1}^{3}+\widehat{c}\mu_{I,1}^{2}+\widehat{d}\mu_{I,1}+\widehat{e}=0\label{eq:relation-between-the-inhibitory-membrane-potentials}
\end{equation}
}
\par\end{center}{\small \par}
\end{spacing}

\noindent whose coefficients depend on $\mu_{I,0}$ as follows:

\begin{spacing}{0.8}
\begin{center}
{\small{}
\begin{align*}
\widehat{a}= & \frac{\Lambda_{I}^{2}}{4\tau_{I}^{2}}\\
\\
\widehat{b}= & -\frac{\Lambda_{I}^{2}}{2\tau_{I}}\left(\widehat{\psi}+\frac{V_{I}^{T}}{\tau_{I}}\right)\\
\\
\widehat{c}= & \frac{\Lambda_{I}^{2}}{4}\left[\widehat{\psi}^{2}+\left(\frac{V_{I}^{T}}{\tau_{I}}\right)^{2}+\frac{4}{\tau_{I}}V_{I}^{T}\widehat{\psi}\right]+\frac{1}{\tau_{I}^{2}}-\widehat{\xi}\\
\\
\widehat{d}= & -\frac{\Lambda_{I}^{2}}{2}\widehat{\psi}V_{I}^{T}\left(\frac{V_{I}^{T}}{\tau_{I}}+\widehat{\psi}\right)-\frac{2}{\tau_{I}}\widehat{\psi}+2\widehat{\xi}V_{I}^{T}\\
\\
\widehat{e}= & \left(\frac{\Lambda_{I}}{2}\widehat{\psi}V_{I}^{T}\right)^{2}+\widehat{\psi}^{2}-\widehat{\xi}\left(V_{I}^{T}\right)^{2}\\
\\
\widehat{\psi}= & \frac{1}{\tau_{I}}\mu_{I,0}+\frac{J_{II}}{N-1}\mathscr{A}_{I}\left(\mu_{I,0}\right)-\frac{\nu_{I}^{\mathrm{max}}J_{II}}{2\left(N-1\right)}\\
\\
\widehat{\xi}= & \left(\frac{\nu_{I}^{\mathrm{max}}\Lambda_{I}J_{II}}{4\left(N-1\right)}\right)^{2}
\end{align*}
}
\par\end{center}{\small \par}
\end{spacing}

\noindent Eq.~\eqref{eq:relation-between-the-inhibitory-membrane-potentials}
can be solved analytically, providing an explicit expression of $\mu_{I,1}$
as a function of $\mu_{I,0}$, which will be used in SubSec.~\eqref{sub:Strong-inhibition-regime}
to evaluate the local bifurcations in the strong-inhibition regime.

\subsection{Weak-inhibition regime ($\lambda_{I}<0$) \label{sub:Weak-inhibition-regime}}

As we said, we want to understand how the network's dynamics changes
when we vary the external input currents $I_{E,I}$ and the strength
of the synaptic weight $J_{II}$. For the sake of clarity, in this
section we show the results that we obtain when we vary these parameters
one by one, because this allows us to introduce the concepts of codimension
one and codimension two bifurcation diagrams.

\begin{figure}
\centering{}\includegraphics[width=0.85\textwidth]{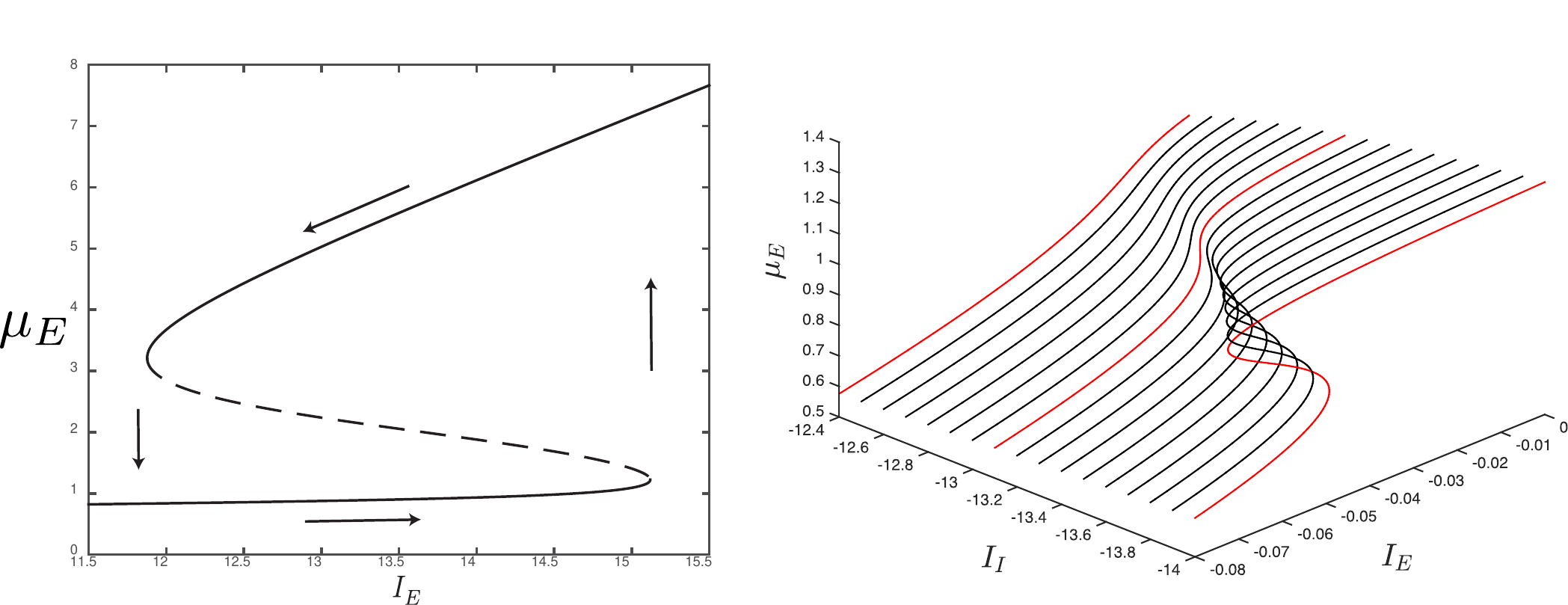} \protect\caption{\label{Fig:catastrophe-manifold} \small On the left, we show an
example of hysteresis displayed by the system. The plain lines describe
stable equilibria, while the dashed line the unstable ones. On the
right, we show an example of catastrophe manifold. The panel highlights
three different behaviors of the network for increasing values of
$\left|I_{I}\right|$: leaky integrator, perfect integrator and switch
(red curves). The readers are referred to \cite{Simen2011,Simen2012}
for more details about their biological relevance.}
\end{figure}

We start by considering fixed $I_{E,I}$ and inhibitory strength $J_{II}=-10$.
As we can see from Fig.~\eqref{Fig:equilibrium-points-2D} (top),
Eq.~\eqref{eq:equilibrium-points-equations-homogeneous-case} admits
multiple solutions, depending on the specific value that we have chosen
for the current $I_{E}$. Then, we start to vary $I_{E}$ continuously,
while keeping $I_{I}$ and $J_{II}$ fixed. In this way we can plot
how the solutions $\mu_{E,I}$ of the system \eqref{eq:equilibrium-points-equations-homogeneous-case}
change as a function of $I_{E}$. As we anticipated in SubSec.~\eqref{sub:Intuitive-interpretation-of-the-branching-points},
the curve that we obtain is called the \textit{primary branch} of
the network, see Fig.~\eqref{Fig:equilibrium-points-2D} (bottom).
This figure represents the \textit{codimension one bifurcation diagram}
of the network, from which we see there are two special points, called
\textit{local saddle-node bifurcations}, or LP for short. They identify
the value of $I_{E}$ for which the number of solutions of Eq.~\eqref{eq:equilibrium-points-equations-homogeneous-case}
changes (notice the correspondence with the top panels of Fig.~\eqref{Fig:equilibrium-points-2D}).
These are the first example of (local) bifurcation points that the
neural network exhibits, and that lead to the formation of hysteresis
(see the left panel in Fig.~\eqref{Fig:catastrophe-manifold}). 
 Hysteresis was suggested to describe short-term memory, since a sufficiently
strong input can lead the system to reach a stable-high level of activity
that is maintained when the input is turned off \cite{Cragg1955,Ingber1984,Wang2001}.
This phenomenon, known as \textit{reverberation}, namely the persistence
of neural activity sustained internally in the brain after a stimulus
is removed, can be achieved through bistability, which can be present
intrinsically at the single cell level or generated by recurrent excitatory
connections. A typical example is the working memory, which is intimately
related to the prefrontal cortex and that can retain information for
a time span of the order of seconds, before it is erased by events
such as noise or distracting stimuli. It is important to observe that
even if reverberation requires bistability, the latter can be present
without hysteresis, but very often they coexist, as in our model.

\begin{figure}
\centering{} \includegraphics[width=0.8\textwidth]{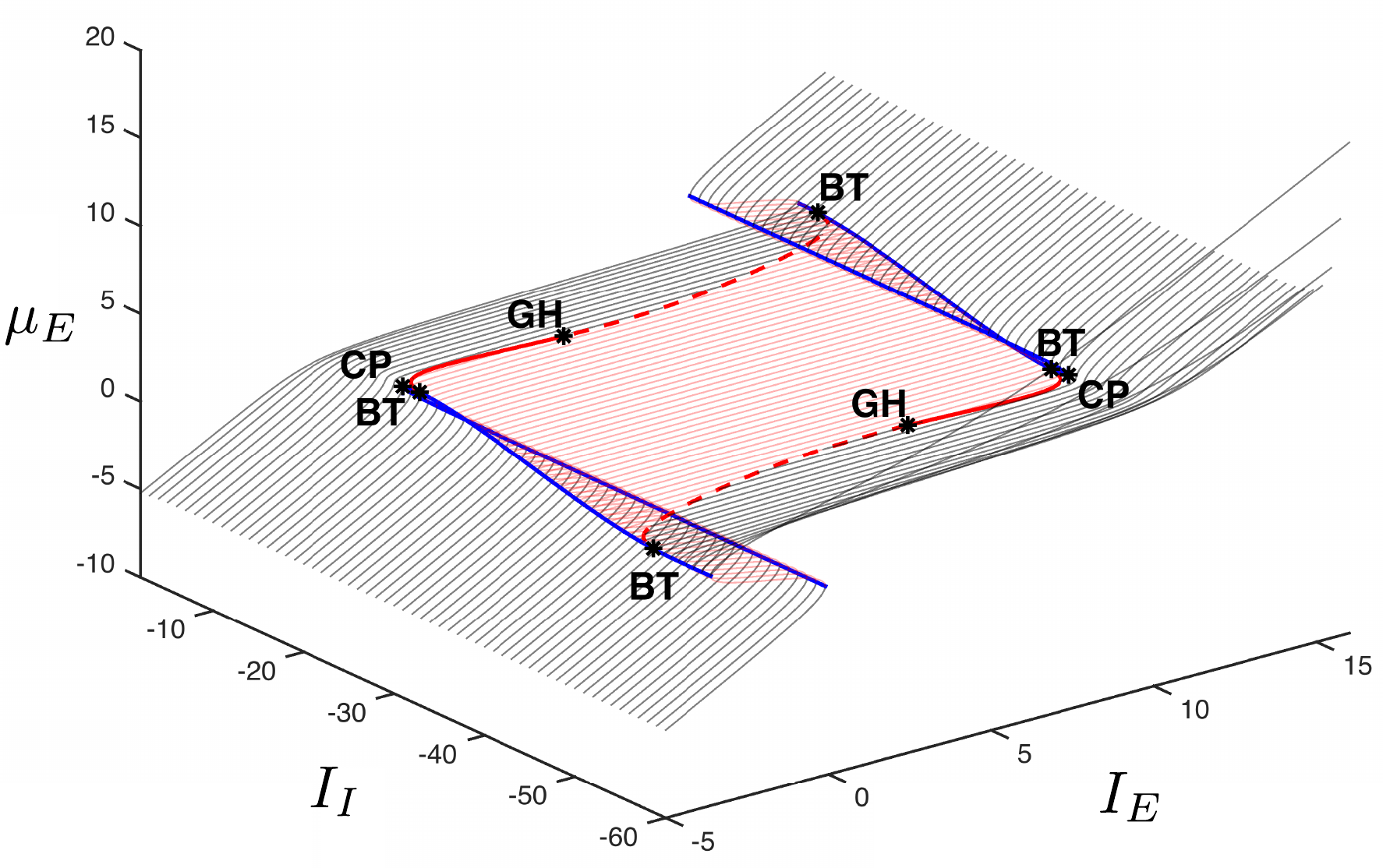} \protect\caption{\label{Fig:codimension-2-bifurcation-diagram-3D} \small Three-dimensional
bifurcation diagram equivalent to that in Fig.~\eqref{Fig:codimension-2-bifurcation-diagram-projection}.
The values of $\mu_{E}$ are shown on the $z-$axis as functions of
$I_{E}-I_{I}$. Here we plot only the local bifurcations (blue: saddle-node
curves, red: Andronov-Hopf curves) that bound the gray/red regions
representing the stable/unstable equilibrium point areas. All the
codimension two bifurcations are reported in Fig.~\eqref{Fig:codimension-2-bifurcation-diagram-projection}.}
\end{figure}

\begin{figure}
\centering{} \includegraphics[scale=0.7]{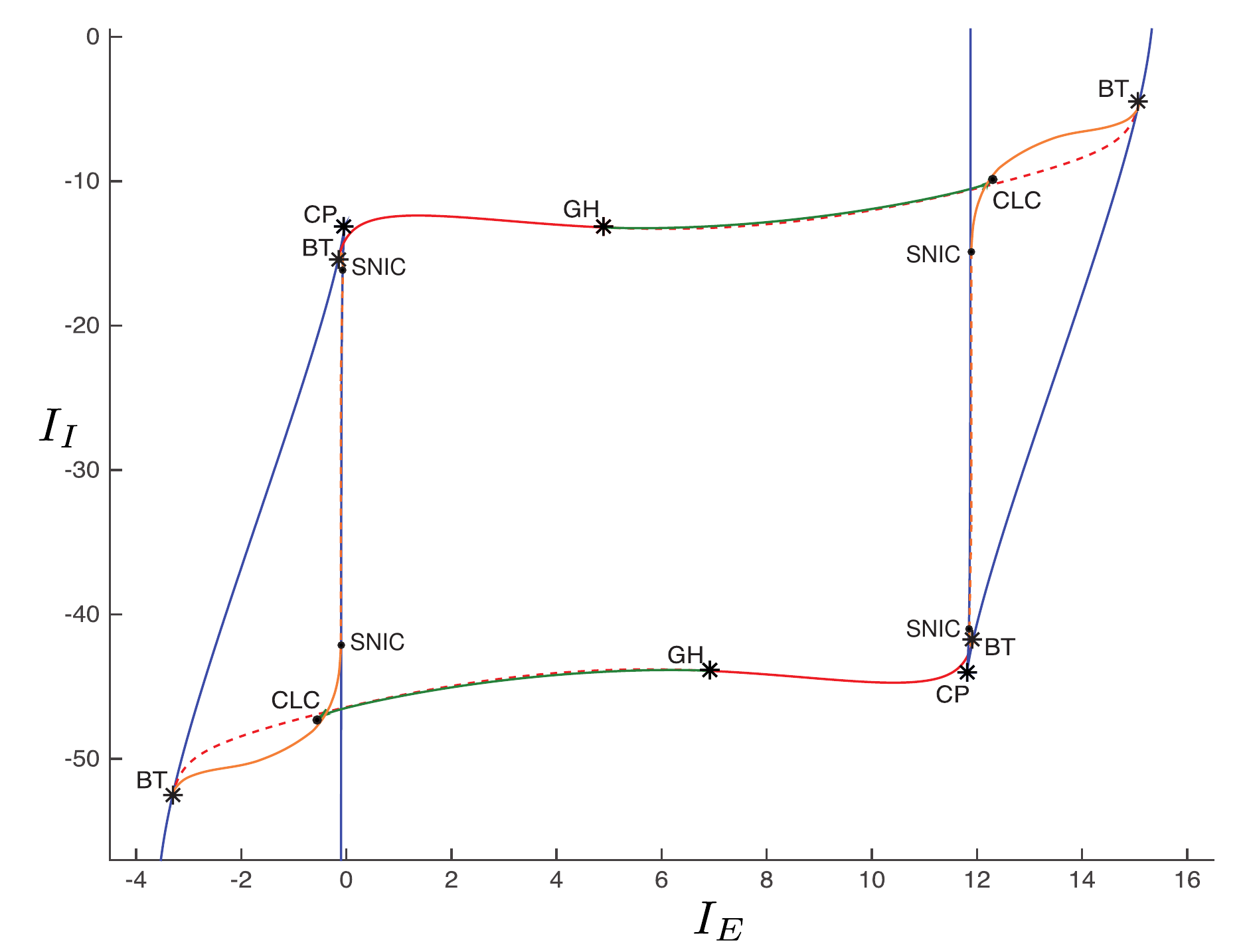} \protect\caption{\label{Fig:codimension-2-bifurcation-diagram-projection} \small
Complete codimension two bifurcation diagram on the $I_{E}-I_{I}$
plane in the weak-inhibition regime. The Andronov-Hopf bifurcation
curves (red lines) are divided into supercritical (plain) and subcritical
(dashed) portions. The supercritical/subcritical portions are bounded
by a Generalized Hopf bifurcation, GH, and Bogdanov-Takens bifurcations,
BT. The latter are the contact points among saddle-node bifurcation
curves (blue lines), Andronov-Hopf bifurcation curves (red lines),
and homoclinic bifurcation curves (hyperbolic-saddle/saddle-node homoclinic
bifurcations are described by plain/dashed orange curves). SNIC bifurcations
identify the contact point between the saddle-node curve and the homoclinic
one. From GH originate two Limit Point of Cycles curves (dark green
lines) that collapse into the homoclinic curves. Before this, they
present a cusp bifurcation, CLC. Each saddle-node curve shows, in
addiction to BT, a cusp bifurcation, CP.}
\end{figure}

If now we continuously vary both $I_{E}$ and $I_{I}$, while keeping
$J_{II}$ fixed, we can plot the solutions of Eq.~\eqref{eq:equilibrium-points-equations-homogeneous-case}
as a function of both the external currents, so now $\mu_{E,I}=\mu_{E,I}\left(I_{E},I_{I}\right)$
define three dimensional manifolds (see Fig.~\eqref{Fig:codimension-2-bifurcation-diagram-3D}
for $\mu_{E}$). 
 On these manifolds, the special points LP depend also on $I_{I}$,
therefore they form a set of points called \textit{saddle-node curves}
(see the blue curves in Fig.~\eqref{Fig:codimension-2-bifurcation-diagram-3D}).
For visual convenience, these curves are projected on the $I_{E}-I_{I}$
plane (see Fig.~\eqref{Fig:codimension-2-bifurcation-diagram-projection}),
defining the so-called \textit{codimension two bifurcation diagram}
of the network (which is, from the analytical point of view, our main
interest in this article, while the codimension one diagram is partially
calculated in Sec.~(S2) of the Supplementary Materials). 
 However, from Fig.~\eqref{Fig:codimension-2-bifurcation-diagram-3D}
and the bottom panels of Fig.~\eqref{Fig:equilibrium-points-2D},
we can see that the saddle-nodes are not the only bifurcations we
have in our network. For some values of the pair $I_{E}-I_{I}$, the
system can undergo also a so called \textit{local Andronov-Hopf bifurcation},
or H for short. These bifurcations are represented by the red curves
in Figs.~\eqref{Fig:codimension-2-bifurcation-diagram-3D} and \eqref{Fig:codimension-2-bifurcation-diagram-projection},
and correspond to the emergence of neural oscillations, which are
thought to play a key role in many cognitive processes \cite{Ward2003}. 

Hereafter, we list all the bifurcations the system undergoes, dividing
them in groups depending on the codimension of the bifurcation, which
is defined as the number of parameters (in our case $I_{E,I}$) that
must be varied for the bifurcation to occur. Although only few of
them are represented in Fig.~\eqref{Fig:equilibrium-points-2D},
the complete set of codimension one bifurcations our system undergoes
is:
\begin{itemize}
\item local saddle-node bifurcations (LP), for which new equilibria arise,
or collide and annihilate each other;
\item local Andronov-Hopf bifurcations (H), where stable or unstable self-sustained
oscillations, described by limit cycles, arise or disappear;
\item global homoclinic bifurcations, where limit cycles vanish in a particular
equilibrium point (i.e. a neutral saddle, see SubSec.~\eqref{sub:Strong-inhibition-regime},
or a saddle-node), giving rise to an orbit with infinite period;
\item global limit point of cycles bifurcations, at which new limit cycles
arise, or collide and annihilate each other.
\end{itemize}
The codimension one diagrams collecting all these bifurcations are
shown in Fig.~(S5) of the Supplementary Materials. Moreover, on the
curves defined by these bifurcations, and that are obtained by varying
both $I_{E}$ and $I_{I}$, the following codimension two bifurcations
appear (see Fig.~\ref{Fig:codimension-2-bifurcation-diagram-projection}):
\begin{itemize}
\item local cusp bifurcations (CP), on the saddle-node curves;
\item local generalized Hopf bifurcations (GH), that divide subcritical
Andronov-Hopf curves from supercritical ones;
\item local Bogdanov-Takens bifurcations (BT), that represent the contact
point between the saddle-node, Andronov-Hopf (ending here) and homoclinic
curves;
\item global cusp of limit point of cycles (CPC), on the limit point of
cycles curves;
\item global saddle-node on invariant circle (SNIC), where a saddle-node
bifurcation occurs simultaneously with a homoclinic bifurcation.
\end{itemize}
It is worth remarking that the bifurcation diagram in Fig.~\eqref{Fig:codimension-2-bifurcation-diagram-projection},
obtained from the voltage-based model \eqref{eq:exact-rate-equation-2}
in the weak-inhibition regime, is qualitatively similar to that of
the activity-based Wilson-Cowan model (see Fig. 2.12 in \cite{Izhikevich1997}).
These two kinds of models are obtained from neural mass equations
through two slightly different hypothesis about the postsynaptic potentials
\cite{Grimbert2008}. For this reason, in the literature it has always
been assumed implicitly that the two models can exhibit qualitatively
similar dynamics. The strong resemblance of their codimension two
bifurcation diagrams proved in this article confirms rigorously this
intuition (notwithstanding, in the next section we will show that
things may change significantly in the strong-inhibition regime if
we take into account finite-size effects).

Interestingly, Fig.~\eqref{Fig:codimension-2-bifurcation-diagram-3D}
presents two of the so-called \textit{catastrophe manifolds} \cite{Zeeman1976},
one of which is shown in the right panel of Fig.~\eqref{Fig:catastrophe-manifold}.
This figure emphasizes the ability of the model to describe three
different behaviors: leaky integrator, perfect integrator and switch.
This triad represents the main ingredient for describing a mechanism
which was proposed to explain interval timing by neural integration
\cite{Simen2011,Simen2012}. According to the theory, by changing
some parameters of the network, it is possible to modify the shape
of the equilibrium curve on the catastrophe manifold, generating a
ramping activity that can explain Weber's law of time perception \cite{Allan1979}.
This phenomenon can easily occur in our model, where the shape of
the equilibrium curve can be changed dynamically by varying the input
currents.

Now we want to prove some of our previous results on bifurcations
from the analytical point of view. Often, in dynamical systems, global
bifurcations are harder to study analytically, therefore here we focus
on LP, H and BT. So first of all we observe that, according to bifurcation
theory \cite{Kuznetsov1998elements,Strogatz1994nonlinear}, LP is
mathematically defined by the condition that one real eigenvalue of
the Jacobian matrix becomes equal to zero. Therefore, from Eq.~\eqref{eq:eigenvalues-Jacobian-matrix},
we conclude that for our network this bifurcation occurs whenever
$\lambda_{0}=0$ or $\lambda_{1}=0$, because $\lambda_{E}$ is always
negative, while $\lambda_{I}=0$ defines the branching point bifurcation.
For example, if $\delta+\eta<0$, the condition $\lambda_{0}=0$ is
equivalent to $\delta\eta=\gamma$ which, according to Eq.~\eqref{eq:parameters-for-the-eigenvalues},
provides:

\begin{spacing}{0.8}
\begin{center}
{\small{}
\begin{equation}
\mathscr{A}_{I}'\left(\mu_{I}\right)=\frac{-\frac{1}{\tau_{E}\tau_{I}}+\frac{1}{\tau_{I}}\frac{N_{E}-1}{N-1}J_{EE}\mathscr{A}_{E}'\left(\mathfrak{v}\right)}{-\frac{1}{\tau_{E}}\frac{N_{I}-1}{N-1}J_{II}+\frac{1}{\left(N-1\right)^{2}}\left[\left(N_{E}-1\right)\left(N_{I}-1\right)J_{EE}J_{II}-N_{E}N_{I}J_{EI}J_{IE}\right]\mathscr{A}_{E}'\left(\mathfrak{v}\right)}\label{eq:derivative-inhibitory-activation-function-saddle-node-points}
\end{equation}
}
\par\end{center}{\small \par}
\end{spacing}

\noindent where we have defined the parameter $\mathfrak{v}\overset{\mathrm{def}}{=}\mu_{E}$.
Now we invert $\mathscr{A}_{I}'\left(\mu_{I}\right)$ (more details
are provided in SubSec.~(S3.2.1) of the Supplementary Materials),
obtaining:

\begin{spacing}{0.8}
\begin{center}
{\small{}
\begin{equation}
\mu_{I}\left(\mathfrak{v}\right)=V_{I}^{T}\pm\frac{2}{\Lambda_{I}}\sqrt{\sqrt[3]{\left(\frac{\nu_{I}^{\mathrm{max}}\Lambda_{I}}{4\mathscr{A}_{I}'\left(\mu_{I}\right)}\right)^{2}}-1}\label{eq:inhibitory-membrane-potential-saddle-node-points}
\end{equation}
}
\par\end{center}{\small \par}
\end{spacing}

\noindent and from Eq.~\eqref{eq:equilibrium-points-equations-homogeneous-case}
we get:

\begin{spacing}{0.8}
\begin{center}
{\small{}
\begin{equation}
\begin{cases}
I_{E}\left(\mathfrak{v}\right)=\frac{1}{\tau_{E}}\mathfrak{v}-\frac{N_{E}-1}{N-1}J_{EE}\mathscr{A}_{E}\left(\mathfrak{v}\right)-\frac{N_{I}}{N-1}J_{EI}\mathscr{A}_{I}\left(\mu_{I}\left(\mathfrak{v}\right)\right)\\
\\
I_{I}\left(\mathfrak{v}\right)=\frac{1}{\tau_{I}}\mu_{I}\left(\mathfrak{v}\right)-\frac{N_{E}}{N-1}J_{IE}\mathscr{A}_{E}\left(\mathfrak{v}\right)-\frac{N_{I}-1}{N-1}J_{II}\mathscr{A}_{I}\left(\mu_{I}\left(\mathfrak{v}\right)\right)
\end{cases}\label{eq:parametric-equations-saddle-node-points}
\end{equation}
}
\par\end{center}{\small \par}
\end{spacing}

\noindent These are parametric equations in the parameter $\mathfrak{v}\in\left(\mathfrak{v}_{a},\mathfrak{v}_{b}\right)$,
where:

\begin{spacing}{0.8}
\begin{center}
{\small{}
\begin{equation}
\mathfrak{v}_{b,a}=V_{E}^{T}\pm\frac{2}{\Lambda_{E}}\sqrt{\sqrt[3]{\left(\frac{N_{E}-1}{N-1}J_{EE}\frac{\nu_{E}^{\mathrm{max}}\Lambda_{E}}{4}\tau_{E}\right)^{2}}-1}\label{eq:extreme-parameters-saddle-node-manifold}
\end{equation}
}
\par\end{center}{\small \par}
\end{spacing}

\noindent and they define analytically the blue curves in Fig.~\eqref{Fig:codimension-2-bifurcation-diagram-projection}.
As we said, this is not sufficient to prove that Eqs.~\eqref{eq:derivative-inhibitory-activation-function-saddle-node-points}
- \eqref{eq:extreme-parameters-saddle-node-manifold} describe saddle-node
curves, since we should check also the corresponding non-degeneracy
conditions. Nevertheless, we observe a perfect agreement between these
analytical curves and those obtained numerically by Cl\_MatCont and
XPPAUT, therefore for simplicity we do not check the remaining conditions
and we leave them to the most technical readers. We adopt the same
approach for the remaining bifurcations we are about to describe.

Now we focus on the H bifurcations. According to \cite{Kuznetsov1998elements,Strogatz1994nonlinear},
they appear whenever the network has conjugate purely imaginary eigenvalues.
Since $\lambda_{E,I}$ are always real, this condition can be satisfied
only by $\lambda_{0,1}$, by setting $\delta+\eta=0$ and $\left(\delta-\eta\right)^{2}+4\gamma<0$.
In particular, from the equation $\delta+\eta=0$ we get:

\begin{spacing}{0.8}
\begin{center}
{\small{}
\begin{equation}
\mathscr{A}_{I}'\left(\mu_{I}\right)=\frac{N-1}{\left(N_{I}-1\right)J_{II}}\left[\frac{1}{\tau_{E}}+\frac{1}{\tau_{I}}-\frac{N_{E}-1}{N-1}J_{EE}\mathscr{A}_{E}'\left(\mathfrak{v}\right)\right]\label{eq:derivative-inhibitory-activation-function-Hopf-points}
\end{equation}
}
\par\end{center}{\small \par}
\end{spacing}

\noindent where $\mathfrak{v}\overset{\mathrm{def}}{=}\mu_{E}$ as
before. Following the same procedure introduced before for the LP
curves, we obtain a set of parametric equations for the pairs $I_{E}-I_{I}$
that generate the H curves, with parameter $\mathfrak{v}\in\left[\mathfrak{v}_{f},\mathfrak{v}_{d}\right]\cup\left[\mathfrak{v}_{c},\mathfrak{v}_{e}\right]$,
where:

\begin{spacing}{0.8}
\begin{center}
{\small{}
\begin{align}
\mathfrak{v}_{c,d}= & V_{E}^{T}\pm\frac{2}{\Lambda_{E}}\sqrt{\sqrt[3]{\left(\frac{\nu_{E}^{\mathrm{max}}\Lambda_{E}\left(N_{E}-1\right)J_{EE}}{4\left(N-1\right)}\frac{1}{\frac{1}{\tau_{E}}+\frac{1}{\tau_{I}}-\frac{\nu_{I}^{\mathrm{max}}\Lambda_{I}\left(N_{I}-1\right)J_{II}}{4\left(N-1\right)}}\right)^{2}}-1}\nonumber \\
\nonumber \\
\mathfrak{v}_{e,f}= & V_{E}^{T}\pm\frac{2}{\Lambda_{E}}\sqrt{\sqrt[3]{\left(\frac{\nu_{E}^{\mathrm{max}}\Lambda_{E}}{4\mathfrak{z}}\right)^{2}}-1}\nonumber \\
\nonumber \\
\mathfrak{z}= & \frac{-\mathfrak{b}-\sqrt{\mathfrak{b}^{2}-4\mathfrak{a}\mathfrak{c}}}{2\mathfrak{a}}\label{eq:extreme-parameters-Hopf-manifold}\\
\nonumber \\
\mathfrak{a}= & \left(\frac{N_{E}-1}{N-1}J_{EE}\right)^{2}-\frac{N_{E}N_{I}\left(N_{E}-1\right)}{\left(N-1\right)^{2}\left(N_{I}-1\right)}\frac{J_{EE}J_{EI}J_{IE}}{J_{II}}\nonumber \\
\nonumber \\
\mathfrak{b}= & -\frac{2}{\tau_{E}}\frac{N_{E}-1}{N-1}J_{EE}+\frac{N_{E}N_{I}}{\left(N-1\right)\left(N_{I}-1\right)}\frac{J_{EI}J_{IE}}{J_{II}}\left(\frac{1}{\tau_{E}}+\frac{1}{\tau_{I}}\right)\nonumber \\
\nonumber \\
\mathfrak{c}= & \frac{1}{\tau_{E}^{2}}\nonumber 
\end{align}
}
\par\end{center}{\small \par}
\end{spacing}

Now, as we said, BT represents the point where the LP and H curves
meet each other, and identifies also the end of the H curve. Clearly,
from the condition $\lambda_{0}=0$ or $\lambda_{1}=0$ that defines
the LP curves, and the condition $\lambda_{0,1}=\pm\iota\omega$ (where
$\iota$ represents the imaginary unit) that defines the H curves,
we get $\lambda_{0}=\lambda_{1}=0$. This is the condition that defines
analytically the BT points, or equivalently $\mathfrak{v}_{\mathrm{BT}}=\mathfrak{v}_{e,f}$
as given by Eq.~\eqref{eq:extreme-parameters-Hopf-manifold}, from
which the coordinates of the BT points in the $I_{E}-I_{I}$ plane
can be easily obtained through Eq.~\eqref{eq:equilibrium-points-equations-homogeneous-case}.
The remaining local bifurcations (i.e. CP and GH) are analytically
intractable, therefore we cannot study them beyond the numerical results.

To conclude this subsection, now we describe briefly the effect of
the variation of the remaining synaptic weights on the codimension
two bifurcation diagram, considering the weights in Tab.~\eqref{Tab:parameters}
as reference point. As we said, their variation does not generate
interesting phenomena such as the branching point bifurcation which
is obtained by varying $J_{II}$, so we dedicate less space to these
parameters.

For $J_{EE}\gg10$, the two LP curves become larger and larger on
the $I_{E}$ axis (i.e. the distance between their vertical asymptotes
increases). Moreover, the curves get closer and closer to each other,
by shifting on the $I_{E}$ axis, until they intersect and their oblique
parts (i.e. those between the BT points) overlap. If we increase $J_{EE}$
further, the LP curves split again in two disjoint parts, each one
presenting two BT and two CP bifurcations (so the total number of
CP points increases from two to four). Between each pair of BT points
(on the same LP curve) there is a H curve. These curves are very close
to the corresponding LP curves, and if we increase $J_{EE}$ further
they disappear, together with the BT bifurcations. So for very large
$J_{EE}$ we get only two disconnected LP curves, or in other terms
for very strong excitation the oscillatory activity cannot be sustained
anymore. Also on the opposite side, namely for weak inhibition (i.e.
$J_{EE}\rightarrow0$), the H curves disappear. Moreover, the width
on the $I_{E}$ axis of the LP curves decreases, i.e. the distance
between their vertical asymptotes becomes smaller and smaller, until
the asymptotes collapse on each other and the LP curves disappear
as well.

For $\left|J_{EI}\right|\gg70$, the width of the two LP curves remains
almost constant, while the distance between them (and therefore also
the length of the H curves) increases continuously. On the other side,
for $\left|J_{EI}\right|\rightarrow0$, the two LP curves get closer
and closer to each other, until they intersect and then their oblique
parts between the BT points overlap. If we decrease $\left|J_{EI}\right|$
further, similarly to the case with large $J_{EE}$, the two LP curves
split in two disjoint parts, each one presenting two BT and two CP
bifurcations. For even smaller values of $\left|J_{EI}\right|$, the
BT, CP and H bifurcations disappear, while the LP curves disappear
for $\left|J_{EI}\right|=0$.

For $J_{IE}\gg70$, the LP curves are stretched vertically and shifted
downwards along the $I_{I}$ axis. Clearly, in the opposite direction
(i.e. for $J_{IE}\rightarrow0$), they are compressed, and while the
$I_{E}$ coordinates of the vertical asymptotes remain almost unchanged
with $J_{IE}$, the two CP points get closer and closer to each other.
At the same time, the two H curves tend to overlap. At some value
of $J_{IE}$, the two CP bifurcations touch each other and disappear,
so the two LP curves become tangent. If we further decrease $J_{IE}$,
the two LP curves split again, one over the other, and the BT and
H bifurcations disappear.

All the phenomena that we have just described are qualitatively similar
for different values of $J_{II}$, so they occur also in the strong-inhibition
regime for the primary branch.

\subsection{Strong-inhibition regime ($\lambda_{I}\geq0$) \label{sub:Strong-inhibition-regime}}

\begin{figure}
\centering{}\includegraphics[scale=0.7]{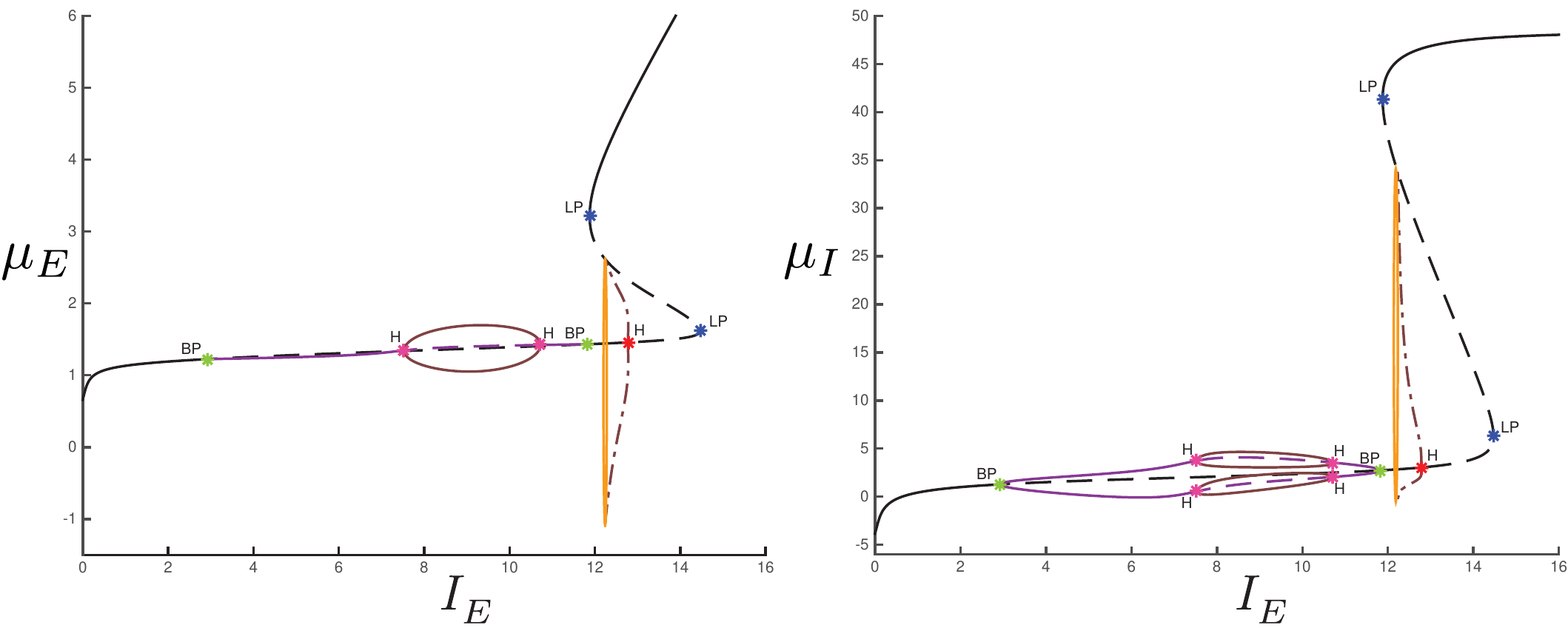} \protect\caption{\label{Fig:strong-inhibition-regime-0} \small Codimension one bifurcation
diagrams for $\mu_{E}$ (left) and $\mu_{I}$ (right) as a function
of $I_{E}$, for $J_{II}=-34$ and $I_{I}=-10$. In both the graphs,
the stable/unstable primary equilibrium curve is described by plain/dashed
black curves, while the secondary ones are described by plain/dashed
violet curves. Andronov-Hopf bifurcation appear on both the primary
and secondary equilibrium curves, giving rise to unstable and stable
limit cycles respectively. In particular, the unstable cycles collapse
into a homoclinic bifurcation, described by an orange line.}
\end{figure}

\begin{figure}
\centering{}\includegraphics[scale=0.72]{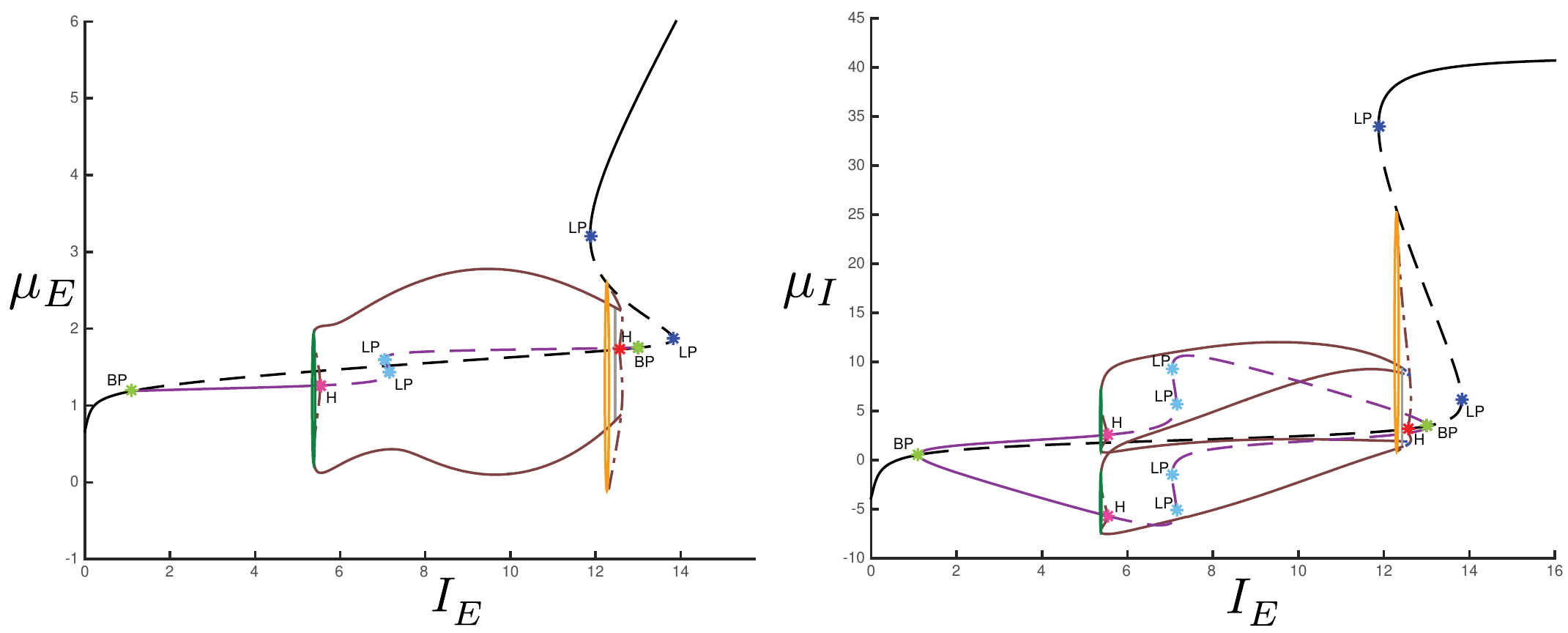} \protect\caption{\label{Fig:strong-inhibition-regime-1} \small Codimension one bifurcation
diagrams for $\mu_{E}$ (left) and $\mu_{I}$ (right) as a function
of $I_{E}$, for $J_{II}=-100$ and $I_{I}=-10$. The colored curves
describe the same bifurcations as in Fig. \eqref{Fig:strong-inhibition-regime-0}.
Besides, we observe an LPC bifurcation (dark green line) and a TR
bifurcation (gray line).}
\end{figure}

\begin{figure}
\centering{}\includegraphics[scale=0.45]{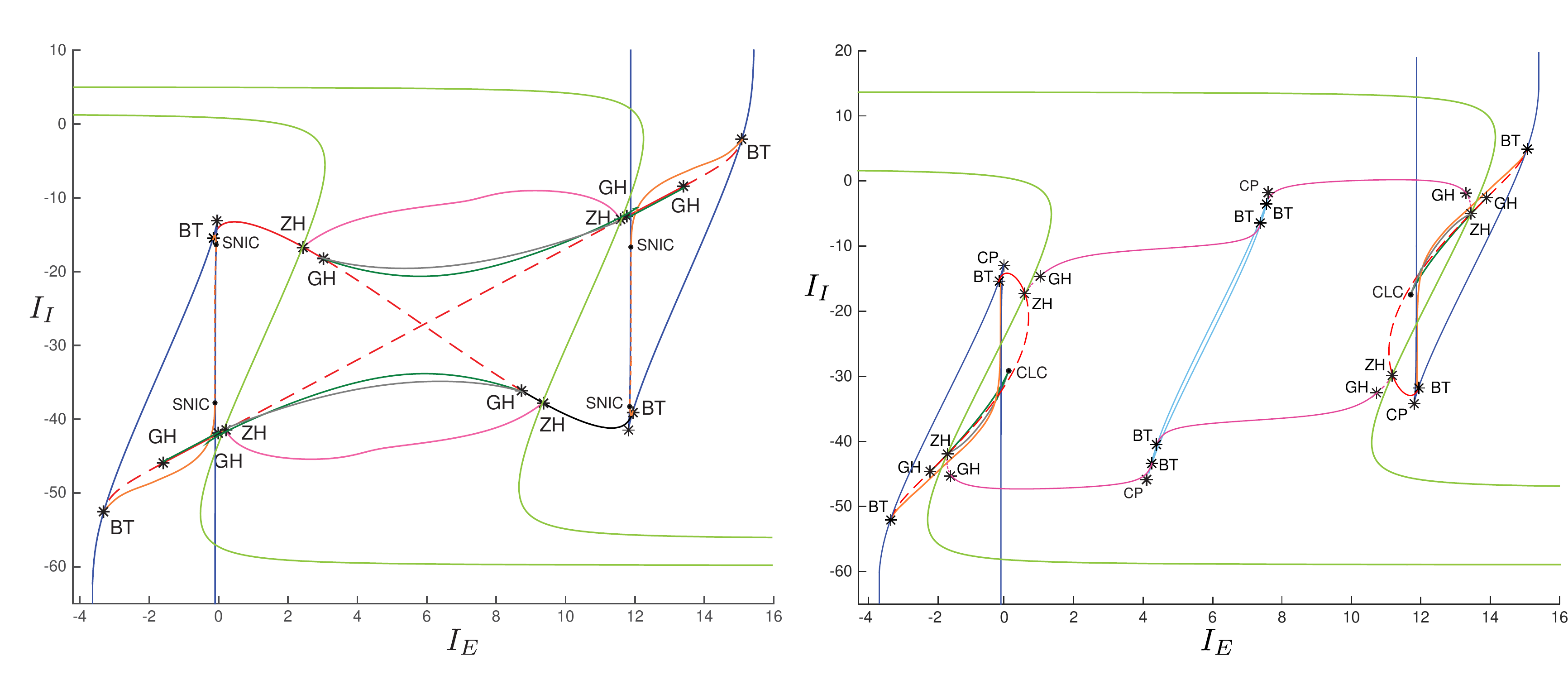} \protect\caption{\label{Fig:strong-inhibition-regime-2} \small Codimension two bifurcation
diagram on the $I_{E}-I_{I}$ plane for $J_{II}=-34$ (left) and $J_{II}=-100$
(right). In addition to the bifurcations already displayed in Fig.~\eqref{Fig:codimension-2-bifurcation-diagram-projection},
we stress the presence of new ones. The branching points form two
curves (light green lines), that define the values of $I_{E}-I_{I}$
that bound the secondary branches of equilibrium points (see the violet
curves in Figs.~\eqref{Fig:strong-inhibition-regime-0} and \eqref{Fig:strong-inhibition-regime-1}).
The bifurcations originated on the secondary branches are differentiated
from those originated on the primary one. Specifically, we show Andronov-Hopf
and saddle-node curves (purple and light blue lines, respectively).
In addition, we display the torus bifurcation curve (gray lines).}
\end{figure}

In the strong-inhibition regime (in particular here we consider the
cases $J_{II}=-34$ and $J_{II}=-100$), most of the features of the
weak-inhibition bifurcation diagram are preserved. However, besides
the bifurcations explained in SubSec.~\eqref{sub:Weak-inhibition-regime},
from Figs.~\eqref{Fig:strong-inhibition-regime-0}, \eqref{Fig:strong-inhibition-regime-1},
\eqref{Fig:strong-inhibition-regime-2}, we can see that the system
undergoes also the following codimension one bifurcations:
\begin{itemize}
\item local branching point bifurcations (BP), at which two or more equilibrium
curves intersect each other;
\item local torus bifurcations (TR), at which the limit cycles are characterized
by a quasi-periodic motion;
\end{itemize}
and the following codimension two bifurcations:
\begin{itemize}
\item local zero-Hopf (neutral saddle) bifurcations (ZH), that involves
the Andronov-Hopf curves and, in our case, the branching point curves.
Around this point, both subcritical and supercritical Andronov-Hopf
bifurcations exist.
\end{itemize}
In particular, the branching point bifurcations lead our model to
show multiple branches of stationary solutions for suitable current
values. This is a finite-size effect, due to the finite number of
neurons in each population, that leads to a richer set of Jacobian
matrix eigenvalues than that obtained by using methods based on the
reduction of the number of equations, such as the mean-field approximation
(see SubSec.~\eqref{sub:Differences-between-our-approach-and-the-mean-field-theory}
for more details). In order to thoroughly investigate the bifurcations
the system undergoes in presence of strong inhibition, we start by
analyzing the codimension one bifurcation diagram for $J_{II}=-34$.
In particular, the diagram in Fig.~\eqref{Fig:strong-inhibition-regime-0}
is obtained by letting $I_{I}=-10$. 
 It turns out that, in addition to the primary equilibrium point curve
(black line), new branches of stationary states (violet lines) emanate
and collapse in two branching point bifurcations, BP. These secondary
branches hold supercritical Andronov-Hopf bifurcations that give rise
to stable limit cycles. Instead, on the primary branch, we find two
saddle-node bifurcations and a subcritical Andronov-Hopf bifurcation,
whose unstable limit cycles vanish in a homoclinic orbit. We remind
that branching point bifurcations occur because $\lambda_{I}$ changes
sign. We also observe that for \emph{$N_{I}=2$ }the inhibitory neurons
have the same bifurcation diagram (see Fig.~\eqref{Fig:strong-inhibition-regime-0},
right). However, this does not mean that the inhibitory membrane potentials
are homogeneous. Indeed, when an inhibitory neuron is on the upper
secondary branch (see the violet curve above the primary equilibrium
point curve in Fig.~\eqref{Fig:strong-inhibition-regime-0}, right),
the other one is in the lower secondary branch, so indeed they are
heterogeneous.

For $J_{II}=-100$, secondary branches of equilibrium points are still
present, see Fig.~\eqref{Fig:strong-inhibition-regime-1}. 
 Together with a subcritical Andronov-Hopf bifurcation, they unveil
also saddle-node ones. In particular, the former generates unstable
limit cycles that become stable after having crossed the limit point
of cycles bifurcation (dark green line). For increasing values of
$I_{E}$, the stable limit cycles collapse into the unstable limit
cycles originated from the subcritical Andronov-Hopf bifurcation belonging
to the primary equilibrium point curve. Before collapsing, the stable
limit cycles undergo torus bifurcations (gray line). 

By varying also $I_{I}$, we obtain the codimension two bifurcation
diagrams displayed in Fig.~\eqref{Fig:strong-inhibition-regime-2}.
 It is worth noting that the branching points the system undergoes
generate two bifurcation curves (light green lines) that pass through
the whole $I_{E}-I_{I}$ domain. The presence of these curves is the
most relevant difference with the weak-inhibition regime and the classic
(mean-field) Wilson-Cowan model. Furthermore, the LP and H bifurcations
that belong to the secondary branches give rise to further bifurcation
curves (purple and light blue lines, respectively) in the $I_{E}-I_{I}$
domain, as shown in Fig.~\eqref{Fig:strong-inhibition-regime-2}. 

Interestingly, since the branching point bifurcation increases the
dimension of the network from $2$ (for $\lambda_{I}<0$, see Eq.~\eqref{eq:equilibrium-points-equations-homogeneous-case})
to $3$ (for $\lambda_{I}>0$, see Eq.~\eqref{eq:equilibrium-points-equations-heterogeneous-case-1}),
the network can exhibit more complex dynamics, such as quasi-periodic
motions originated from the torus bifurcations. The biological importance
of quasi-periodic oscillations in neural communication was discussed
in \cite{Izhikevich1999}.

Now we want to study the local bifurcations from the analytical point
of view. We start by considering the BP bifurcations, which are defined
by the condition $\lambda_{I}=0$, as we saw before. From Eq.~\eqref{eq:eigenvalues-Jacobian-matrix}
this condition implies:

\begin{spacing}{0.8}
\noindent \begin{center}
{\small{}
\begin{equation}
\mathscr{A}_{I}'\left(\mu_{I}\left(\mathrm{BP}\right)\right)=\frac{N-1}{\tau_{I}\left|J_{II}\right|}\label{eq:condition-branching-point-bifurcations}
\end{equation}
}
\par\end{center}{\small \par}
\end{spacing}

\noindent so the solutions of this equation are:

\begin{spacing}{0.8}
\begin{center}
{\small{}
\begin{equation}
\mu_{I}\left(\mathrm{BP}\right)=V_{I}^{T}\pm\frac{2}{\Lambda_{I}}\sqrt{\sqrt[3]{\left(\frac{\tau_{I}\left|J_{II}\right|\nu_{I}^{\mathrm{max}}\Lambda_{I}}{4\left(N-1\right)}\right)^{2}}-1}\label{eq:inhibitory-membrane-potential-branching-points}
\end{equation}
}
\par\end{center}{\small \par}
\end{spacing}

\noindent Now, from the second equation of the system \eqref{eq:equilibrium-points-equations-homogeneous-case}
(we can also use Eq.~\eqref{eq:equilibrium-points-equations-heterogeneous-case-1},
since for $\lambda_{I}=0$ they are equivalent) we get:

\begin{spacing}{0.8}
\begin{center}
{\small{}
\begin{equation}
\mu_{E}\left(\mathrm{BP}\right)=V_{E}^{T}\pm\frac{2}{\Lambda_{E}}\sqrt{\frac{1}{\frac{1}{\left\{ \frac{2\left(N-1\right)}{\nu_{E}^{\mathrm{max}}N_{E}J_{IE}}\left[\frac{1}{\tau_{I}}\mu_{I}\left(\mathrm{BP}\right)-\frac{N_{I}-1}{N-1}J_{II}\mathscr{A}_{I}\left(\mu_{I}\left(\mathrm{BP}\right)\right)-I_{I}\right]-1\right\} ^{2}}-1}}\label{eq:excitatory-membrane-potential-branching-points}
\end{equation}
}
\par\end{center}{\small \par}
\end{spacing}

\noindent while from the first equation of \eqref{eq:equilibrium-points-equations-homogeneous-case}
we get:

\begin{spacing}{0.8}
\begin{center}
{\small{}
\begin{equation}
I_{E}=\frac{1}{\tau_{E}}\mu_{E}\left(\mathrm{BP}\right)-\frac{N_{E}-1}{N-1}J_{EE}\mathscr{A}_{E}\left(\mu_{E}\left(\mathrm{BP}\right)\right)-\frac{N_{I}}{N-1}J_{EI}\mathscr{A}_{I}\left(\mu_{I}\left(\mathrm{BP}\right)\right)\label{eq:excitatory-current-branching-points}
\end{equation}
}
\par\end{center}{\small \par}
\end{spacing}

\noindent where $\mu_{I}\left(\mathrm{BP}\right)$ and $\mu_{E}\left(\mathrm{BP}\right)$
are given by Eqs.~\eqref{eq:inhibitory-membrane-potential-branching-points}
and \eqref{eq:excitatory-membrane-potential-branching-points} respectively.
Since $\mu_{E}\left(\mathrm{BP}\right)$ depends on $I_{I}$, Eq.~\eqref{eq:excitatory-current-branching-points}
defines two explicit functions $I_{E}=\mathcal{F}_{\pm}\left(I_{I}\right)$,
that provide the curves on which we have a branching point bifurcation
(see the light green lines in Fig.~\eqref{Fig:strong-inhibition-regime-2}
for $J_{II}=-34$ and $J_{II}=-100$. More details can be found in
SubSec.~(S4.2.1) of the Supplementary Materials).

The points where the H and BP curves meet each other define the ZH
bifurcation. From this definition, we see that they can be calculated
analytically from the conditions $\lambda_{0,1}=\pm\iota\omega$ and
$\lambda_{I}=0$, from which in turn we get:

\begin{spacing}{0.8}
\begin{center}
{\small{}
\begin{align*}
\mathscr{A}_{E}'\left(\mu_{E}\left(\mathrm{ZH}\right)\right)= & \frac{N-1}{\left(N_{E}-1\right)J_{EE}}\left(\frac{1}{\tau_{E}}+\frac{N_{I}}{\tau_{I}}\right)\\
\\
\mathscr{A}_{I}'\left(\mu_{I}\left(\mathrm{ZH}\right)\right)= & \frac{N-1}{\tau_{I}\left|J_{II}\right|}
\end{align*}
}
\par\end{center}{\small \par}
\end{spacing}

\noindent and therefore:

\begin{spacing}{0.8}
\begin{center}
{\small{}
\begin{align*}
\mu_{E}^{\pm}\left(\mathrm{ZH}\right)= & V_{E}^{T}\pm\frac{2}{\Lambda_{E}}\sqrt{\sqrt[3]{\left(\frac{\nu_{E}^{\mathrm{max}}\Lambda_{E}\left(N_{E}-1\right)J_{EE}}{4\left(N-1\right)\left(\frac{1}{\tau_{E}}+\frac{N_{I}}{\tau_{I}}\right)}\right)^{2}}-1}\\
\\
\mu_{I}^{\pm}\left(\mathrm{ZH}\right)= & V_{I}^{T}\pm\frac{2}{\Lambda_{I}}\sqrt{\sqrt[3]{\left(\frac{\nu_{I}^{\mathrm{max}}\Lambda_{I}\tau_{I}\left|J_{II}\right|}{4\left(N-1\right)}\right)^{2}}-1}
\end{align*}
}
\par\end{center}{\small \par}
\end{spacing}

\noindent As usual, if we substitute these expressions of the membrane
potentials in Eq.~\eqref{eq:equilibrium-points-equations-homogeneous-case}
or \eqref{eq:equilibrium-points-equations-heterogeneous-case-1},
we obtain the coordinates of the ZH points in the $I_{E}-I_{I}$ plane.

As we said, on the secondary branches that are generated by the branching
points, new bifurcations can occur (in the case $N_{I}=2$, see for
example the LP and H bifurcations in Figs.~\eqref{Fig:strong-inhibition-regime-0}
and \eqref{Fig:strong-inhibition-regime-1}, and the corresponding
light blue and purple curves in Fig.~\eqref{Fig:strong-inhibition-regime-2}),
also new branching points (for $N_{I}>2$), from which tertiary branches
emerge, and so on. To study them, according to bifurcation theory,
we need the Jacobian matrix of the network on the secondary (tertiary,
and so on) branches, as we will explain more clearly in Sec.~\eqref{sub:The-case-with-generic-NI}.
As usual, we focus on the case $N_{I}=2$, therefore we can determine
the local bifurcations on the secondary branches by means of Eq.~\eqref{eq:Jacobian-matrix-on-the-new-branches}.

Now we start with the LP bifurcations. We know that they are defined
by the condition that one of the eigenvalues of \eqref{eq:Jacobian-matrix-on-the-new-branches}
is equal to zero. From it, as explained in SubSec.~(S4.2.3) of the
Supplementary Materials, we obtain that:

\begin{spacing}{0.8}
\begin{center}
{\small{}
\begin{equation}
\mathscr{A}_{E}'\left(\mu_{E}\right)=\frac{\acute{b}}{\acute{a}}\label{eq:derivative-exctitatory-activation-function-saddle-node-points-secondary-branches}
\end{equation}
}
\par\end{center}{\small \par}
\end{spacing}

\noindent where:

\begin{spacing}{0.8}
\begin{center}
{\small{}
\begin{align*}
\acute{a}= & \frac{1}{\tau_{I}^{2}}\frac{N_{E}-1}{N-1}J_{EE}+\frac{1}{\tau_{I}}\frac{N_{E}}{\left(N-1\right)^{2}}J_{EI}J_{IE}\left[\mathscr{A}_{I}'\left(\mu_{I,0}\right)+\mathscr{A}_{I}'\left(\mu_{I,1}\right)\right]\\
 & +\frac{1}{\left(N-1\right)^{3}}\left[2N_{E}J_{EI}J_{IE}J_{II}-\left(N_{E}-1\right)J_{EE}J_{II}^{2}\right]\mathscr{A}_{I}'\left(\mu_{I,0}\right)\mathscr{A}_{I}'\left(\mu_{I,1}\right)\\
\\
\acute{b}= & \frac{1}{\tau_{E}}\left[\frac{1}{\tau_{I}^{2}}-\left(\frac{J_{II}}{N-1}\right)^{2}\mathscr{A}_{I}'\left(\mu_{I,0}\right)\mathscr{A}_{I}'\left(\mu_{I,1}\right)\right]
\end{align*}
}
\par\end{center}{\small \par}
\end{spacing}

\noindent So if we invert Eq.~\eqref{eq:derivative-exctitatory-activation-function-saddle-node-points-secondary-branches}
and we use the solution of Eq.~\eqref{eq:relation-between-the-inhibitory-membrane-potentials},
we obtain the expression of $\mu_{E}$ as a function of $\mu_{I,0}$.
If we replace the solutions $\mu_{E}$ and $\mu_{I,1}$ in the system
\eqref{eq:equilibrium-points-equations-heterogeneous-case-1}, we
get parametric equations for $I_{E,I}$ as a function of a single
parameter, which is now defined as $\mathfrak{v}\overset{\mathrm{def}}{=}\mu_{I,0}$.
These equations are an analytical description of the light blue curves
shown in Fig.~\eqref{Fig:strong-inhibition-regime-2} (right) for
$J_{II}=-100$.

Similarly, for the H bifurcations we obtain the condition:

\begin{spacing}{0.8}
\begin{center}
{\small{}
\begin{equation}
\mathscr{A}_{E}'\left(\mu_{E}^{\pm}\right)=\frac{-\grave{b}\pm\sqrt{\grave{b}^{2}-4\grave{a}\grave{c}}}{2\grave{a}}\label{eq:derivative-exctitatory-activation-function-Hopf-points-secondary-branches}
\end{equation}
}
\par\end{center}{\small \par}
\end{spacing}

\noindent where:

\begin{spacing}{0.8}
\begin{center}
{\footnotesize{}
\begin{align*}
\grave{a}= & \frac{N_{E}-1}{N-1}J_{EE}\left[\frac{2}{\tau_{I}}\frac{N_{E}-1}{N-1}J_{EE}+\frac{N_{E}}{\left(N-1\right)^{2}}J_{EI}J_{IE}\left(\mathscr{A}_{I}'\left(\mu_{I,0}\right)+\mathscr{A}_{I}'\left(\mu_{I,1}\right)\right)\right]\\
\\
\grave{b}= & 2\frac{N_{E}}{\left(N-1\right)^{3}}J_{EI}J_{IE}J_{II}\mathscr{A}_{I}'\left(\mu_{I,0}\right)\mathscr{A}_{I}'\left(\mu_{I,1}\right)-\left(\frac{1}{\tau_{E}}+\frac{1}{\tau_{I}}\right)\left[\frac{4}{\tau_{I}}\frac{N_{E}-1}{N-1}J_{EE}+\frac{N_{E}}{\left(N-1\right)^{2}}J_{EI}J_{IE}\left(\mathscr{A}_{I}'\left(\mu_{I,0}\right)+\mathscr{A}_{I}'\left(\mu_{I,1}\right)\right)\right]\\
\\
\grave{c}= & \frac{2}{\tau_{I}}\left[\left(\frac{1}{\tau_{E}}+\frac{1}{\tau_{I}}\right)^{2}-\left(\frac{J_{II}}{N-1}\right)^{2}\mathscr{A}_{I}'\left(\mu_{I,0}\right)\mathscr{A}_{I}'\left(\mu_{I,1}\right)\right]
\end{align*}
}
\par\end{center}{\footnotesize \par}
\end{spacing}

\noindent so again it is possible to describe these bifurcations analytically,
obtaining the same results we got numerically in Fig.~\eqref{Fig:strong-inhibition-regime-2}
for $J_{II}=-34$ and $J_{II}=-100$ (see the purple curves in both
the panels). However, unlike the primary branch, our theory does not
allow us to calculate the range of the parameter $\mathfrak{v}=\mu_{I,0}$
on the secondary branches, since the resulting equations that define
the range are analytically intractable. In the same way, now it is
not possible to calculate explicitly the coordinates of the new BT
bifurcations, where the LP and H curves that emanate from the secondary
branches meet each other. Therefore analytical approximations or numerical
schemes must be used to evaluate them, but this is beyond the purpose
of the article.

As we did for the weak-inhibition regime, we conclude this section
by describing briefly the effect of the variation of the remaining
synaptic weights. As we said at the end of SubSec.~\eqref{sub:Weak-inhibition-regime},
all the variations that occur on the primary branch are qualitatively
similar for different values of $J_{II}$. On the other side, now
we want to analyze the behavior of the BP curves and of the bifurcations
on the secondary branches. For $J_{II}=-100$ and increasing $J_{EE}$,
the most notable phenomenon is the overlapping between the oblique
parts of the BP and the LP curves of the primary branch. The latter
finally collapse on each other and split in two disjoint parts as
in the case $J_{II}=-10$ discussed in SubSec.~\eqref{sub:Weak-inhibition-regime},
while the bifurcations on the secondary branches do not show any interesting
variation. Furthermore, when $J_{EE}\rightarrow0$, we observe first
of all the disappearance of the ZH bifurcations. This occurs because
the H curves on both the primary and the secondary branches do not
meet the BP curve anymore. If we further decrease $J_{EE}$, the two
CP bifurcations on the LP curve of the secondary branches get closer
and closer until they annihilate each other and the curve disappears.
Clearly this phenomenon implies also the disappearance of the BT bifurcations
on the secondary branches. For smaller values of $J_{EE}$, the H
curves on both the primary and secondary branches disappear, and finally
also the LP curve on the primary branch (see SubSec.~\eqref{sub:Weak-inhibition-regime})
and the BP curve. To conclude, for large $\left|J_{EI}\right|$ or
large $J_{IE}$, the LP curve on the secondary branches disappears
again through the annihilation of its CP points (as explained above),
while on the other side, when at least one of the two synaptic weights
is small, we do not observe any interesting variation of the bifurcations
on the secondary branches.

\subsection{The case with generic $N_{I}$ \label{sub:The-case-with-generic-NI}}

The same analysis can be performed on networks with a generic number
$N_{I}$ of inhibitory neurons. When $\lambda_{I}$, as given by Eq.~\eqref{eq:eigenvalues-Jacobian-matrix},
goes to zero, we observe in general the formation of secondary branches
from the primary one. This means that an inhibitory membrane potential
becomes different from the others, so we can reinterpret the system
as a network with an excitatory population with $N_{E}$ neurons,
and two inhibitory populations, one with one neuron, and the other
with $N_{I}-1$ neurons. Furthermore, when we change the current $I_{E}$
(while keeping $\lambda_{I}>0$) with $I_{I}$ fixed, at some point
one of the eigenvalues of the Jacobian matrix of the system \eqref{eq:equilibrium-points-equations-heterogeneous-case-0}
(see the Thm.~(S1) of the Supplementary Materials for their analytical
calculation) may go to zero, generating a new branching point on the
secondary branches. In this case, we observe the formation of tertiary
branches, so now one of the previous $N_{I}-1$ identical inhibitory
membrane potentials becomes different from the others, and so on.

\begin{figure}
\centering{}\includegraphics[scale=0.5]{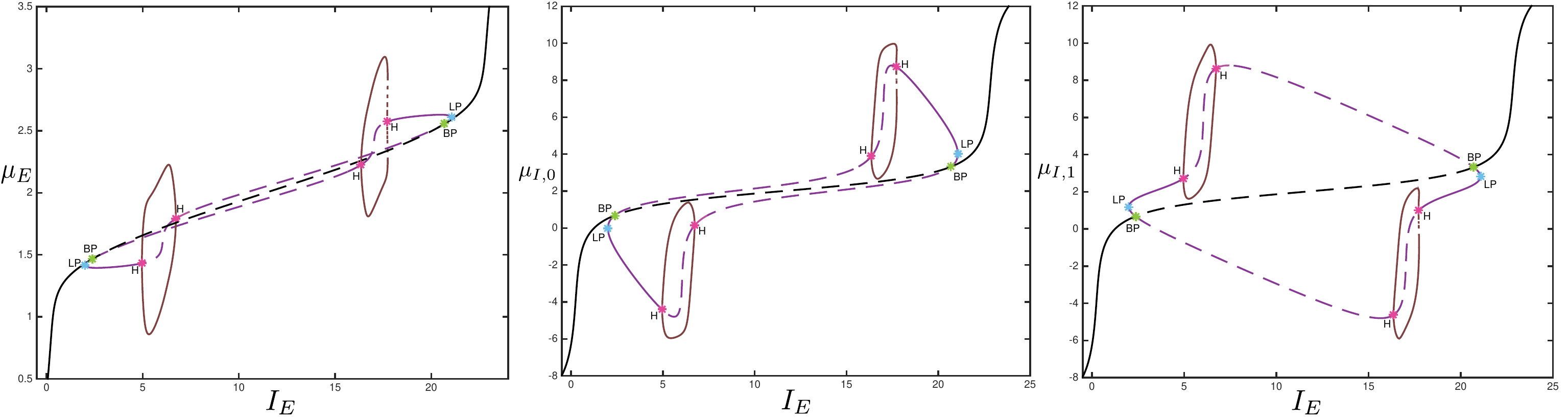}\protect\caption{\label{Fig:strong-inhibition-regime-3} \small Codimension one bifurcation
diagrams for $N_{I}=4$. The left panel shows the diagram of the excitatory
neurons, the central panel that of three of the inhibitory neurons,
while the right panel shows the diagram of the remaining neuron in
the inhibitory population. Interestingly, now the inhibitory neurons
have not only different membrane potentials, but also different codimension
one bifurcation diagrams. The neuron with the different diagram is
chosen randomly by the system, so this is another example of symmetry-breaking
that occurs in the network for strong inhibition.}
\end{figure}

In Fig.~\eqref{Fig:strong-inhibition-regime-3} we show an example
of formation of secondary branches in the case $N_{I}=4$, obtained
as usual for the values of the parameters reported in Tab.~\eqref{Tab:parameters}.
 As the reader may easily see, there is now an important difference
compared to the case $N_{I}=2$. For a network with only two inhibitory
neurons, Eq.~\eqref{eq:equilibrium-points-equations-heterogeneous-case-1}
implies that they both have the same codimension one bifurcation diagram
(see the right panels of Figs.~\eqref{Fig:strong-inhibition-regime-0}
and \eqref{Fig:strong-inhibition-regime-1}). This is just a special
case, because in general, for $N_{I}>2$, we observe a symmetry-breaking
not only at the level of the inhibitory membrane potentials, but also
at the level of the codimension one bifurcation diagram. Indeed, from
Fig.~\eqref{Fig:strong-inhibition-regime-3} we see that one inhibitory
neuron, which is chosen randomly by the system, has a different diagram
compared to the others. This is another example of symmetry-breaking
that occurs in the system, as an implicit consequence of Eq.~\eqref{eq:equilibrium-points-equations-heterogeneous-case-0}
for $N_{I}>2$.

It is important to underline that, even if the BP curve \eqref{eq:excitatory-current-branching-points}
is defined for every $N_{I}$, numerically we observe the formation
of new branches only for $N_{I}$ even. In principle this may be proved
rigorously by the so called \textit{Lyapunov-Schmidt bifurcation equation}
\cite{Sidorov2002}, but since the proof is rather technical and beyond
the purpose of this article, we do not report it here.

For a generic $N_{I}$, local bifurcations can still be calculated
analytically as we showed before. Indeed, from the second equation
of \eqref{eq:equilibrium-points-equations-heterogeneous-case-0},
it is possible to express $N_{I}-1$ inhibitory membrane potentials
as functions of the remaining one, which can be used as a parameter
for the parametric equations in the codimension two bifurcation diagram.

To conclude, we observe that now new kinds of bifurcations can appear,
which do not occur in the case $N_{I}=2$. For example, for $N_{I}=4$,
if the network has four different inhibitory potentials, the characteristic
equation of the Jacobian matrix has the form $p\left(\lambda\right)=\left(\lambda-\lambda_{E}\right)^{N_{E}-1}p_{R}\left(\lambda\right)$,
where $p_{R}\left(\lambda\right)$ (the characteristic polynomial
of the reduced Jacobian matrix introduced in Thm.~(S1) of the Supplementary
Materials) is a fifth order polynomial. This means that in principle,
for some values of the parameters, the network may have two pairs
of purely imaginary eigenvalues. This condition corresponds to the
formation of a so called \textsl{double-Hopf bifurcation}, which in
turn may imply a local birth of \textit{chaos} (e.g. \cite{Bi1999,Yu2002}).
Indeed, for $\lambda_{I}>0$ the dimension of the system is larger
than $2$ due to the BP bifurcations, therefore according to the Poincaré\textendash Bendixson
theorem the network may exhibit chaotic behavior.

\subsection{Differences between our approach and the mean-field theory \label{sub:Differences-between-our-approach-and-the-mean-field-theory}}

In this section we explain in detail why the mean-field theory cannot
describe the branching point bifurcation. Given the system \eqref{eq:exact-rate-equation-2},
the mean-field theory makes the assumption that within each population
the membrane potentials are independent and identically distributed.
Therefore, by hypothesis, it forbids the presence of heterogeneous
solutions, like those that emerge from the branching point bifurcation.
Due to this assumption, Eq.~\eqref{eq:exact-rate-equation-2} can
be reduced to a system of two differential equations, according to
the mean-field theory developed by Sznitman, Tanaka, McKean and others
\cite{Sznitman1984,Sznitman1986,Sznitman1991,Tanaka1978,Tanaka1981,Tanaka1983,McKean1966,McKean1967}:

\begin{spacing}{0.80000000000000004}
\begin{center}
{\small{}
\begin{equation}
\begin{cases}
\frac{dV_{E}\left(t\right)}{dt}= & \frac{1}{\tau_{E}}V_{E}\left(t\right)+R_{E}J_{EE}\mathbb{E}\left[\mathscr{A}_{E}\left(V_{E}\left(t\right)\right)\right]+R_{I}J_{EI}\mathbb{E}\left[\mathscr{A}_{I}\left(V_{I}\left(t\right)\right)\right]+I_{E}\\
\\
\frac{dV_{I}\left(t\right)}{dt}= & \frac{1}{\tau_{I}}V_{I}\left(t\right)+R_{E}J_{IE}\mathbb{E}\left[\mathscr{A}_{E}\left(V_{E}\left(t\right)\right)\right]+R_{I}J_{II}\mathbb{E}\left[\mathscr{A}_{I}\left(V_{I}\left(t\right)\right)\right]+I_{I}
\end{cases}\label{eq:mean-field-equations}
\end{equation}
}
\par\end{center}{\small \par}
\end{spacing}

\noindent where $R_{\alpha}=\underset{N\rightarrow\infty}{\lim}\frac{N_{\alpha}}{N}$
(namely the ratio between the population $\alpha$ and the whole network
in the thermodynamic limit, so in our case $R_{E}=0.8$ and $R_{I}=0.2$),
while $V_{\alpha}$ represents any membrane potential in the population
$\alpha$. Moreover, $\mathbb{E}\left[\cdot\right]$ denotes the average
over trials at a given time instant, and it means that the system
is generally supposed to be stochastic. Stochasticity can be introduced
in different ways, for example through Brownian motions, random initial
conditions, or random synaptic weights \cite{Fasoli2015}. In this
article we are considering a deterministic system, therefore we get
simply $\mathbb{E}\left[\mathscr{A}_{\alpha}\left(V_{\alpha}\left(t\right)\right)\right]=\mathscr{A}_{\alpha}\left(V_{\alpha}\left(t\right)\right)$.
In this way, the neural network is described by a system of two coupled
equations in the unknowns $V_{E,I}\left(t\right)$, whose Jacobian
matrix is:

\begin{spacing}{0.8}
\begin{center}
{\small{}
\begin{equation}
\mathcal{J}^{\mathrm{mf}}=\left[\begin{array}{cc}
-\frac{1}{\tau_{E}}+R_{E}J_{EE}\mathscr{A}_{E}'\left(\mu_{E}\right) & R_{I}J_{EI}\mathscr{A}_{I}'\left(\mu_{I}\right)\\
R_{E}J_{IE}\mathscr{A}_{E}'\left(\mu_{E}\right) & -\frac{1}{\tau_{I}}+R_{I}J_{II}\mathscr{A}_{I}'\left(\mu_{I}\right)
\end{array}\right]\label{eq:Jacobian-matrix-mean-field}
\end{equation}
}
\par\end{center}{\small \par}
\end{spacing}

\noindent Its characteristic equation is:

\begin{spacing}{0.80000000000000004}
\begin{center}
{\small{}
\[
a^{\mathrm{mf}}\left(\lambda_{0,1}^{\mathrm{mf}}\right)^{2}+b^{\mathrm{mf}}\lambda_{0,1}^{\mathrm{mf}}+c^{\mathrm{mf}}=0
\]
}
\par\end{center}{\small \par}
\end{spacing}

\noindent where:

\begin{spacing}{0.8}
\begin{center}
{\small{}
\begin{align*}
a^{\mathrm{mf}}= & 1\\
\\
b^{\mathrm{mf}}= & \frac{1}{\tau_{E}}+\frac{1}{\tau_{I}}-R_{E}J_{EE}\mathscr{A}_{E}'\left(\mu_{E}\right)-R_{I}J_{II}\mathscr{A}_{I}'\left(\mu_{I}\right)\\
\\
c^{\mathrm{mf}}= & \frac{1}{\tau_{E}\tau_{I}}-\left(\frac{1}{\tau_{E}}R_{I}J_{II}\mathscr{A}_{I}'\left(\mu_{I}\right)+\frac{1}{\tau_{I}}R_{E}J_{EE}\mathscr{A}_{E}'\left(\mu_{E}\right)\right)+R_{E}R_{I}\left(J_{EE}J_{II}-J_{IE}J_{EI}\right)\mathscr{A}_{E}'\left(\mu_{E}\right)\mathscr{A}_{I}'\left(\mu_{I}\right)
\end{align*}
}
\par\end{center}{\small \par}
\end{spacing}

\noindent From Eq.~\eqref{eq:eigenvalues-Jacobian-matrix} it easy
to see that $\underset{N\rightarrow\infty}{\lim}\lambda_{0,1}=\lambda_{0,1}^{\mathrm{mf}}$.
The only difference between $\lambda_{0,1}$ and $\lambda_{0,1}^{\mathrm{mf}}$
is in the ratios that multiply the synaptic weights ($\frac{N_{\alpha}-1}{N-1}$
or $\frac{N_{\alpha}}{N-1}$ for the finite-size network, and $R_{\alpha}$
in the mean-field case). This difference, due to the absence of self
connections, is small for large networks. So in a sense, when compared
to a finite-size network, the mean-field approximation takes into
account only the information provided by $\lambda_{0,1}$, and neglects
that of $\lambda_{E,I}$. Clearly $\lambda_{E}$ is always negative,
therefore it never affects the changes of dynamics of the system.
However, in a finite-size network $\lambda_{I}$ can change sign,
generating a branching point bifurcation. The mean-field approximation
neglects this information, and it can be seen as a consequence of
the fact that $\underset{N\rightarrow\infty}{\lim}\lambda_{I}=-\frac{1}{\tau_{I}}$.
In other words, in the thermodynamic limit the eigenvalue $\lambda_{I}$
is always negative, therefore it cannot generate branching point bifurcations.
In more mathematically rigorous terms, we get that the \textit{center
manifold} \cite{Kuznetsov1998elements} of the network is not affected
anymore by $\lambda_{I}$ for $N\rightarrow\infty$, so that the dynamics
is governed only by $\lambda_{0,1}$. This clearly proves that the
mean-field approximation oversimplifies the description of the network,
since it is able to describe only the primary branch.

\subsection{Finite-size effects are stronger for biologically plausible anatomical
connections \label{sub:Finite-size-effects-are-stronger-for-biologically-plausible-anatomical-connections}}

\noindent By comparing the right panels of Figs.~\eqref{Fig:strong-inhibition-regime-0}
and \eqref{Fig:strong-inhibition-regime-1}, it is clear that the
magnitude of the branching point finite-size effects is proportional
to the magnitude of $J_{II}$. Indeed, from these figures we can see
that, for a given $I_{E}$, the difference between the membrane potentials
of the primary and secondary branches increases with $\left|J_{II}\right|$.
Moreover, from Fig.~\eqref{Fig:strong-inhibition-regime-2} we observe
that for $J_{II}=-100$ the codimension two diagram is more complex
than the case $J_{II}=-34$, due to the formation of LP, CP and BT
bifurcations on the secondary branches. On the other side, in SubSec.~\eqref{sub:Differences-between-our-approach-and-the-mean-field-theory}
we have seen that the branching point bifurcations disappear in the
thermodynamic limit. This is a consequence of the increasing number
of incoming connections $M_{I}=N-1$, which implies that $\underset{N\rightarrow\infty}{\lim}\lambda_{I}=-\frac{1}{\tau_{I}}<0$.
Therefore from all these observations we conclude that the magnitude
of the finite size effects is related to the ratio $\frac{J_{II}}{M_{I}}$
in the formula of $\lambda_{I}$ (see Eq.~\eqref{eq:eigenvalues-Jacobian-matrix}).
In other terms, both inhibition and the topology of the anatomical
connections determine the strength of the finite-size effects. In
particular, here we want to discuss the importance of the parameter
$M_{I}$.

For this reason, we extend our analysis to the case of a more realistic
connectivity matrix (to simplify matters, we consider a purely inhibitory
neural network, since it is sufficient to show branching point bifurcations).
For example, we can consider the block-circulant topology $\mathcal{BC}_{F,G}\left(\mathcal{M}_{0},\ldots,\mathcal{M}_{F-1}\right)$
with circulant-band blocks introduced in \cite{Fasoli2015}:

\begin{spacing}{0.80000000000000004}
\begin{center}
{\small{}
\begin{align}
J= & J_{II}\left[\begin{array}{cccc}
\mathfrak{B}^{\left(0\right)} & \mathfrak{B}^{\left(1\right)} & \cdots & \mathfrak{B}^{\left(F-1\right)}\\
\mathfrak{B}^{\left(F-1\right)} & \mathfrak{B}^{\left(0\right)} & \cdots & \mathfrak{B}^{\left(F-2\right)}\\
\vdots & \vdots & \ddots & \vdots\\
\mathfrak{B}^{\left(1\right)} & \mathfrak{B}^{\left(2\right)} & \cdots & \mathfrak{B}^{\left(0\right)}
\end{array}\right],\nonumber \\
\label{eq:symmetric-circulant-band-matrix}\\
\mathfrak{B}^{\left(i\right)}= & \left[\begin{array}{cccccccccc}
1-\delta_{i0} & 1 & \cdots & 1 & 0 & \cdots & 0 & 1 & \cdots & 1\\
1 & 1-\delta_{i0} & \ddots &  & \ddots & \ddots &  & \ddots & \ddots & \vdots\\
\vdots & \ddots & \ddots & \ddots &  & \ddots & \ddots &  & \ddots & 1\\
1 &  & \ddots & \ddots & \ddots &  & \ddots & \ddots &  & 0\\
0 & \ddots &  & \ddots & \ddots & \ddots &  & \ddots & \ddots & \vdots\\
\vdots &  & \ddots &  & \ddots & \ddots & \ddots &  & \ddots & 0\\
0 &  &  & \ddots &  & \ddots & \ddots & \ddots &  & 1\\
1 & \ddots &  &  & \ddots &  & \ddots & \ddots & \ddots & \vdots\\
\vdots & \ddots & \ddots &  &  & \ddots &  & \ddots & 1-\delta_{i0} & 1\\
1 & \cdots & 1 & 0 & \cdots & 0 & 1 & \cdots & 1 & 1-\delta_{i0}
\end{array}\right]\nonumber 
\end{align}
}
\par\end{center}{\small \par}
\end{spacing}

\noindent where $\mathfrak{B}^{\left(0\right)},...,\mathfrak{B}^{\left(F-1\right)}$
are $G\times G$ circulant matrices (so that $FG=N$), with bandwidth
$2\xi_{i}+1$, for $i=0,...,F-1$. The network equipped with these
synaptic connections can be interpreted as a collection of $F$ neural
masses with $G$ neurons each. If we define:

\begin{spacing}{0.8}
\begin{center}
{\small{}
\[
H\left(x\right)=\begin{cases}
0, & \begin{array}{cc}
 & x\leq0\end{array}\\
\\
1, & \begin{array}{cc}
 & x>0\end{array}
\end{cases}
\]
}
\par\end{center}{\small \par}
\end{spacing}

\noindent then $\mathcal{M}_{0}\overset{\mathrm{def}}{=}2\xi_{0}-H\left(\xi_{0}-\left\lfloor \frac{G}{2}\right\rfloor +\left(-1\right)^{G}\right)$
is the number of connections that every neuron in a given mass receives
from the neurons in the same mass. Furthermore, $\mathcal{M}_{i}\overset{\mathrm{def}}{=}2\xi_{i}+1-H\left(\xi_{i}-\left\lfloor \frac{G}{2}\right\rfloor +\left(-1\right)^{G}\right)$,
for $i=1,...,F-1$, is the number of connections that every neuron
in the $j$th mass receives from the neurons in the $\left(i+j\right)$th
$\mathrm{mod}\:F$ mass, for $j=0,...,F-1$. Therefore we conclude
that $M_{I}=F-1+\sum_{i=0}^{F-1}\left[2\xi_{i}-H\left(\xi_{i}-\left\lfloor \frac{G}{2}\right\rfloor +\left(-1\right)^{G}\right)\right]$.

Now, if we suppose that the membrane potentials are homogeneous, the
eigenvalues of the corresponding Jacobian matrix are:

\begin{spacing}{0.80000000000000004}
\begin{center}
{\small{}
\begin{align}
\lambda_{mG+n}= & \begin{cases}
-\frac{1}{\tau}+\frac{J_{II}}{M_{I}}\left[F-1+{\displaystyle \sum_{i=0}^{F-1}}g\left(n,\xi_{i},G\right)\right]\mathscr{A}_{I}'\left(\mu_{I}\right), & \begin{array}{cc}
 & m=0,\,\forall n\end{array}\\
\\
-\frac{1}{\tau}+\frac{J_{II}}{M_{I}}\left[-1+{\displaystyle \sum_{i=0}^{F-1}}e^{\frac{2\pi}{F}mi\iota}g\left(n,\xi_{i},G\right)\right]\mathscr{A}_{I}'\left(\mu_{I}\right), & \begin{array}{cc}
 & m\neq0,\,\forall n\end{array}
\end{cases}\nonumber \\
\label{eq:eigenvalues-symmetric-circulant-band-matrix}\\
g\left(n,\xi_{i},G\right)= & \begin{cases}
2\xi_{i}-H\left(\xi_{i}-\left\lfloor \frac{G}{2}\right\rfloor +\left(-1\right)^{G}\right), & \begin{array}{cc}
 & n=0,\:\forall\xi_{i}\end{array}\\
\\
-1, & \begin{array}{cc}
 & n\neq0,\:\xi_{i}=\left\lfloor \frac{G}{2}\right\rfloor \end{array}\\
\\
\frac{\sin\left(\frac{\pi n\left(2\xi_{i}+1\right)}{G}\right)}{\sin\left(\frac{\pi n}{G}\right)}-1, & \begin{array}{cc}
 & n\neq0,\:\xi_{i}<\left\lfloor \frac{G}{2}\right\rfloor \end{array}
\end{cases}\nonumber 
\end{align}
}
\par\end{center}{\small \par}
\end{spacing}

\noindent with $m=0,...,F-1$ and $n=0,...,G-1$. They depend on the
ratio $\frac{J_{II}}{M_{I}}$, where $M_{I}$ does not necessarily
diverge in the thermodynamic limit (consider for example the case
when $G\rightarrow\infty$ for $F$ fixed, and the parameters $\xi_{i}$
are finite and independent from $G$). Therefore, for $N$ large enough,
this topology exhibits stronger finite-size effects than the fully
connected network.

To conclude, we also underline that, according to \cite{Fasoli2015},
if $M_{I}$ does not diverge for $N\rightarrow\infty$, the neurons
do not become independent, therefore Sznitman's mean-field theory
cannot be used to simplify the description of the network. Moreover,
from \eqref{eq:eigenvalues-symmetric-circulant-band-matrix} we see
that many of the eigenvalues $\lambda_{mG+n}$ are distinct, since
the reduced number of connections breaks the degeneracy of the system
(compared to the fully-connected network, where $\lambda_{I}$ has
algebraic multiplicity $N_{I}-1$). For this reason we argue that
they generate a multitude of branching point bifurcations, not just
two as in the fully-connected case, making plausible connections even
more interesting from the biological and mathematical point of view.

\section{Discussion\label{sec:Discussion}}

\noindent We proved the emergence of complex dynamics in small neural
circuits, characterized by strong finite-size effects, that cannot
be accounted for by the mean-field approximation. We showed, through
a detailed numerical and analytical analysis of the bifurcations,
that small symmetric neural networks undergo branching point bifurcations
through spontaneous symmetry-breaking, that leads to the formation
of strongly heterogeneous membrane potentials in the inhibitory population.
This result is obtained when we increase the strength of the synaptic
weights in the inhibitory population, and clearly falsifies the mean-field
hypothesis of identically distributed neurons. This is a very interesting
feature of our model, since from the simple assumption of identical
neurons, it is able to exhibit heterogeneous membrane potentials,
as in real networks.

From a biological point of view, strong inhibition may correspond
to anesthetized neurons, since it was proved that some kinds of anesthetics,
such as propofol, thiopental and isoflurane, act on $\gamma$-Aminobutyric
acid (GABA), the primary inhibitory neurotransmitter in the brain
\cite{Garcia2010}. Our results prove that the dynamics repertoire
of neural populations can be completely different when the brain is
under the influence of drugs, with important consequences on experiments
with anesthetized animals. They also underline the importance of the
synaptic weights in determining the dynamical behavior of the neural
network. As in the theory of hallucinations of Ermentrout and Cowan
\cite{Ermentrout1979}, the spontaneous symmetry-breaking is caused
by drugs that increase the strength of the synaptic weights and generate
alternative neural patterns. However, it is important to observe that
while in their theory the network undergoes a symmetry-breaking through
an increase of the excitatory weights, in our model symmetry is broken
by an increase of the inhibitory ones.

The analysis we performed can be used to understand not only the dynamics
of neural masses at the mesoscopic scale in big animals, but also
the behavior of tiny brains such as those of rotifers and nematodes.
However, it is important to observe that in this article we restricted
the bifurcation analysis to small neural circuits, because of our
hypothesis of all-to-all synaptic connectivity between neurons. This
hypothesis allowed us to reduce the complexity of the analytical formulas,
but predicted that the magnitude of the finite size-effects is inversely
proportional to the number of incoming connections per neuron, that
for a fully connected topology is proportional to the network's size.
Therefore, after relaxing the hypothesis of all-to-all connectivity,
in order to study the behavior of biologically more plausible networks,
we argued the formation of multiple branching point bifurcations and
a stronger heterogeneity of the membrane potentials, as a consequence
of the reduced number of synaptic connections. This means that in
principle strong finite-size effects can occur also in large networks,
if their anatomical connections are sufficiently sparse.

We also observe that our study may be extended to gain some insights
into the oscillations that emerge from Hopf bifurcations. In the codimension
one bifurcation diagram, when the current $I_{E}$ is close to the
one that generates a Hopf bifurcation, the amplitude of the oscillation
that appears from the H point is small. For this reason, the solution
of the system \eqref{eq:exact-rate-equation-2} can be approximated
by a truncated Fourier series. Semi-analytical expressions of the
amplitudes of the harmonics can be obtained by applying the so called
\textit{harmonic balance methods}, which were developed especially
in the context of electrical engineering \cite{Mees1979}. In principle,
these tools may be used to study the formation of torus bifurcations
and eventually the transition to chaos \cite{Genesio1992,Tesi1996,Itovich2005,Itovich2006}.

Furthermore, we stress the fact that the analytical study of local
bifurcations introduced in our work can be easily extended to neural
networks composed of several populations, by means of the Thm.~(S1)
reported in the Supplementary Materials. In this way we can shed new
light on the behavior of more complex and biologically plausible networks,
also when numerical simulations become prohibitive. So for example
we may think to extend our model to describe the interaction between
six neural populations. These can be thought as a coarse description
of the six layers of the cerebral cortex, therefore this system can
be interpreted a simplified model of a cortical column. In general
we obtain that in a network with $\mathfrak{P}$ populations, a codimension
$\mathscr{C}$ bifurcation is described by a $\left(\mathfrak{P}-\mathscr{C}\right)$-dimensional
manifold, whose parametric equations contain $\mathfrak{P}-\mathscr{C}$
independent parameters. The only exception, as in the case with two
populations considered in this article, is the BP manifold, whose
equation can be written by expressing any current as an explicit function
of all the others (see Eqs.~\eqref{eq:inhibitory-membrane-potential-branching-points}
+ \eqref{eq:excitatory-membrane-potential-branching-points} + \eqref{eq:excitatory-current-branching-points}).
So for example, in a network with three populations $A$, $B$, $C$,
in the codimension three diagram spanned by the currents $I_{A}$,
$I_{B}$, $I_{C}$ we get that:
\begin{itemize}
\item the bifurcations LP, H etc are described by surfaces with two parameters;
\item BT, CP etc by lines with one parameter;
\item codimension three bifurcations by points;
\item the BP bifurcations are described by surfaces with explicit formulas
$I_{A}=\mathcal{F}\left(I_{B},I_{C}\right)$.
\end{itemize}
However, it is important to observe that a complete classification
of bifurcations with $\mathscr{C}\geq3$ is still missing in the literature,
due to their complexity. Nevertheless the method introduced in this
article in principle can be used to describe analytically some of
them for any $\mathscr{C}$.

We underline that the Thm.~(S1) can be used to describe also networks
of interconnected neural masses (where each one is described by the
system in Fig.~\ref{Fig:network-structure}). Interestingly, in the
special case when the masses are identical, the symmetry group is
more complex than that of a single mass. As we proved, we may observe
spontaneous symmetry-breaking \textit{within} a mass, through branching
point bifurcations. However, since a network of masses has a larger
symmetry, now symmetry-breaking can occur also \textit{between} masses,
which means that masses can behave differently even if they are identical.
We found that in each mass the inhibitory membrane potentials can
oscillate in synchrony or out of synchrony (on the primary and secondary
branches, respectively). So in principle we may observe the formation
of particular patterns of activity known as \textit{chimera states}
\cite{Abrams2004}, where inhibitory neurons in one mass are synchronized
while those in the other are not, even if the two masses are described
by identical neural equations.

To conclude, we observe that our model can be extended and generalized
in many other ways, through the introduction of delays, stochasticity,
synaptic plasticity and more realistic topologies. All these add-ons
allow us to improve the biological plausibility of the model, without
losing the possibility to investigate analytically the complexity
of its dynamics.

\noindent \textbf{
}

\section*{Acknowledgments}

DF and AC were supported by the Autonomous Province of Trento, Call
\textquotedblleft Grandi Progetti 2012,\textquotedblright{} project
\textquotedblleft Characterizing and improving brain mechanisms of
attention\textemdash ATTEND''.

\noindent \begin{flushleft}
SP was supported by the SI-CODE project of the Future and Emerging
Technologies (FET) programme within the Seventh Framework Programme
for Research of the European Commission, under FET-Open grant no.
FP7-284553, and by the Autonomous Province of Trento, Call \textquotedblleft Grandi
Progetti 2012,\textquotedblright{} project \textquotedblleft Characterizing
and improving brain mechanisms of attention\textemdash ATTEND''.
\par\end{flushleft}

\noindent We would like to thank Fabio Della Rossa from the department
of Electronics, Information and Bioengineering of Politecnico di Milano
(Italy), for providing some useful hints about Cl\_MatCont.

\noindent \begin{flushleft}
The funders had no role in study design, data collection and analysis,
decision to publish, interpretation of results, or preparation of
the manuscript.
\par\end{flushleft}

\end{document}